\documentclass{jfm}
\usepackage{graphicx}
\usepackage{epstopdf, epsfig}

\usepackage{subcaption} 
\usepackage{amsmath}
\usepackage{epstopdf} 
\usepackage[retainorgcmds]{IEEEtrantools}
\usepackage[colorlinks=true,linkcolor=black,citecolor=blue]{hyperref}

\newcommand{\ud}{\mathrm{d}\,}
\newcommand{\e}{\mathrm{e}}

\renewcommand{\i}{\mathrm{i}}
\newcommand{\add}[1]{{\color{black}#1}}
\newcommand{\bxi}{{\color{black}\xi}}
\graphicspath{{Figs/}{IllustrationFigs/}{DataFigs/}}
\shorttitle{Scattering Green's function for a flat plate with a serrated edge}
\shortauthor{B. Lyu}

\title{Analytical Green's function for the acoustic scattering by a flat plate
with \add{a serrated edge}}

\author{B. Lyu\aff{1}
    \corresp{\email{b.lyu@pku.edu.cn}} 
}

\affiliation{\aff{1}State Key Laboratory for Turbulence and Complex
    Systems, College of Engineering, Peking University, 
Beijing 100871, China}
\begin{document}
\maketitle 
\begin{abstract}     
    An analytical Green’s function is developed to study the acoustic
    scattering by a flat plate with \add{a serrated edge}. The scattered
    pressure is solved using the Wiener-Hopf technique in conjunction with the
    adjoint technique. It is shown that the kernel decomposition proposed in
    recent literature appears only valid at high frequencies. We focus on this
    high-frequency regime and obtain the scattered pressure in the form of a
    contour integral. We show that such an integral, although complicated, can
    be evaluated exactly for any arbitrary piecewise linear serrations, for
    which closed-form analytical Green's functions are obtained. The derivation
    is validated by performing numerical integration of the contour integral
    showing excellent agreement. The Green's function is shown to agree well
    with the numerical results obtained using the finite element method at high
    frequencies. The noise directivity patterns are studied as a function of
    the frequency, serration amplitude, source position and Mach number
    respectively. It is found that noise is often enhanced at low and may be
    slightly reduced at high observer angles, which may be understood from the
    perspective of an extended or removed rigid reflection surface. It is found
    that increasing the mean-flow Mach number leads to increasingly evident
    noise amplification at side angles, a seemingly strange Doppler behaviour
    exhibited in source-fixed coordinate frames. The analytical Green's
    function is applicable to both leading- and trailing-edge scatterings, and
    is particularly suitable for developing a three-dimensional trailing-edge
    noise model that is not only highly efficient but also capable of including
    non-frozen turbulence effects.  
\end{abstract}
\section{Introduction}
\label{sec:Introduction}
Turbulent boundary layer trailing-edge (TE) noise~\citep{Howe1978} refers to
the noise generated when turbulence boundary layers convect past the trailing
edge of an aerofoil. It is a common aeroacoustic source in many applications
involving rotating blades such as wind turbines. It has gained increasingly
more attention in recent years, particularly in the wind industry. This is
because the turbine blade size continues to increase, leading to increasingly
large  blade tip velocity. It is well known that the power of the aeroacoustic
noise emission increases quickly as the blade velocity increases, and for
modern wind turbines TE noise has become the dominant noise
source~\citep{Oerlemans2009}. As noise regulations become increasingly
stringent, TE noise is also expected to become a regulatory issue for emerging
commercial transport such as air taxis, small aerial vehicles and
drones~\citep{Jaworksi2020}. Understanding TE noise and its reduction is of
particular importance in these areas. 

There have been numerous studies into the techniques of reducing TE noise. Some
notable approaches include using porous
aerofoils~\citep{Howe1979,Fink1980,Geyer2009a,Geyer2010}, trailing-edge
brushes~\citep{Herr2005}, surface finlets~\citep{Clark2016,Clark2017} and TE
serrations, among which TE serrations represent a particularly effective way of
reducing TE noise without severely compromising aerodynamic efficiency. The
idea of using serrations was inspired by the silent flight of
owls~\citep{Jaworksi2020}. \cite{Thorpe1962} represents one of the earliest
attempts to measure the aeroacoustic signature of free-flying owls. It was
found that the noise generated by owls could not be detected by their
experimental rig in the ultrasonic frequency range. The noise generated by owls
was significantly weaker than that by other birds of similar sizes,
demonstrating the owl's silent flight capability. Later experimental studies by
\cite{Kroeger1972} and \cite{Neuhaus1973} confirmed that owls did have a unique
flying signature that is quieter than other birds. Consistent fly-over noise
measurements by \cite{Sarradj2011}, in conjunction with fixed-wing laboratory
measurement~\citep{Geyer2013}, showed that the silent flying characteristics of
owls may be related to the special features of their wings. One of these
features is the wavy or serrated features around the wing's leading and
trailing edge. \add{The leading-edge serrations appear to be able to reduce the
tip-vortex strength at high angles of attack, whereas TE serrations reduce the
TE noise in approach/gliding flight.} This inspires the technique of installing
serrations on the leading and trailing edges of a wing or blade to reduce its
aerodynamic noise.

Extensive research into TE noise suppression using serrations has been
conducted in the past two decades. Numerous experiments show that serrations
represent an effective technique to reduce TE noise. \citet{Dassen1996}
conducted wind tunnel measurements to study the noise reduction effects of
serrations on aerofoils and flat plates. It was shown that \add{significant
noise reductions} can be achieved in both cases, for example, noise reductions
up to $8$ dB and $10$ dB were observed for aerofoils and flat plates,
respectively. \add{Maximal noise reductions were shown to occur between $1-6$
kHz}. \citet{Parchen1999} undertook a similar experimental campaign, but on
wind turbine blades at both full and laboratory scales. \add{Similar noise
reduction was observed, while noise increase was reported in the high frequency
regime when serrations were misaligned with the flow direction.} A decade
later, \citet{Oerlemans2009} conducted field acoustic measurements on
full-scale wind turbine blades using standard, optimized and serrated blades,
respectively. It was shown that the optimized and serrated blades resulted in a
noise reduction of $0.5$dB and $3.2$dB respectively for a microphone array
placed on the ground. It was found that most of the noise was produced during
the downwash movement of the blades. Gruber {\em et al.}
\citep{Gruber2010,Gruber2012} performed an extensive array of measurements to
study the noise reduction effects of serrations of varying sizes. The Sound
Power Level (SWL) was obtained by integrating noise intensity along a
microphone arc placed in the mid-span plane. An average reduction of $3$-$5$ dB
was reported by using sharp sawtooth serrations. The noise reduction was found
to be related to the change of convection velocity and turbulence coherence
near the serrations. The serrations used by \citet{Gruber2012} are flat
inserts, \citet{Chong2013a} on the other hand studied non-flat serrations by
directly cutting aerofoils and found similar noise reductions. However,
significant boundary layer instability tones were also observed in some
configurations. Recently, \citet{Leon2017} studied the effects of serrations
under deflected configurations. It was found that when the serration was
aligned with the flow a consistent noise reduction up to $7$ dB was obtained,
whereas when the serration was misaligned noise increase started to appear
beyond a critical Strouhal number that scaled with the boundary layer thickness
and freestream velocity. The noise reduction characteristics of serrations when
used specifically on flat plates were studied by \citet{Moreau2013} and
\citet{Chong2015}. Effective noise reduction was reported in both studies. For
example, a noise reduction up to $13$ dB was recorded by \citet{Moreau2013},
but this was shown to be due to the attenuation of vortex shedding. In
\citet{Chong2015}, it was found that little change in the power spectral
density and spanwise correlation length of the surface pressure fluctuations
occurred. Instead, a pair of pressure-driven oblique vortical structures was
identified by using conditional-averaging techniques. In recent years,
experiments were conducted to explore the optimal serration shapes, including
for example serrations with double wavelength ~\citep{Chaitanya2018b},
iron-shaped serrations~\citep{Avallone2017}, ogee serrations~\citep{Lyu2018b}
etc. More details about these experiments can be found in recent
studies~\citep{Lyu2018b}.

In addition to experiments, numerical simulations are also used to study
serrated TE noise. For example, Jones~\citep{Jones2012} performed a Direct
Numerical Simulation (DNS) of flows around a NACA0012 aerofoil with and without
serrations. The serrations appeared to introduce little change into the
turbulent boundary properties and an effective noise reduction was observed.
\citet{Sanjose2014} also performed a DNS on a serrated isolated aerofoil and
reported a noise reduction of a similar magnitude. Numerical studies were also
performed using the Lattice Boltzmann method by \citet{Avallone2018}, where the
link between the far-field noise and the near-field flow parameters was
proposed. In addition to the noise reduction obtained by using conventional
sawtooth serrations, it was shown that combed-sawtooth trailing edges can
provide additional noise reduction benefits.

Both experiments and numerical simulations show that TE serration is indeed an
effective method of reducing TE noise. To use serrations in practical
applications, however, reliable noise predictions models are essential because
they are crucial in the design of optimal serration geometries (see for example
a recent study by \citet{Kholodov2021}). Howe~\citep{Howe1991a,Howe1991b} is
among the earliest researchers to model the aerodynamic noise generated by
serrated trailing edges analytically. A tailored Green's function was used to
formulate a noise prediction model using the blocked surface pressure
statistics beneath the turbulent boundary layers. However, it was well reported
that Howe's model significantly overpredicted the noise reduction by using
serrations. Later studies~\citep{Lyu2015,Lyu2016a} show that this is due to the
Green's function being inaccurate. In order to improve the accuracy of TE noise
prediction, \citet{Lyu2016a} developed a TE noise model using Amiet's approach.
Instead of using the Green's function, the Schwartszchild technique was used in
conjunction with Fourier expansion in an iterative manner to enable analytical
progression. The resulting prediction model yielded more realistic predictions
compared to Howe's model and showed that noise reduction is achieved mainly
through a destructive interference mechanism. The computation of the model
involves the evaluation of nested sums, therefore needs to be optimized so as
to be more suitable when used for serration optimization purposes. Recently
\citet{Ayton2018d} developed a model using the Wiener-Hopf technique. The
far-field sound was formulated as two infinite sums and one infinite integral,
therefore consuming significant time when evaluated. However, it was
shown~\citep{Lyu2019d} that the model can be further developed by evaluating
the infinite integral and one of the infinite sums explicitly, and the
resulting simplified model can be computed very efficiently~\citep{Lyu2019d}.
However, the model hinges on the semi-infinite flat plate assumption and the
result is therefore strictly two dimensional. As such, the far-field pressure
varies as $1/\sqrt{r}$ instead of $1/r$ as $r\to \infty$, where $r$ denotes the
radial distance of the observer in the plane perpendicular to the spanwise
axis. When compared quantitatively with experimental data, it is unclear how
far the microphone should be placed to the serration so that both the
two-dimensionality and far-field assumption are valid simultaneously. More
importantly, since most practical applications involve rotating blades, where
three-dimensionality is crucial, a 3D accurate model would be necessary in
order to obtain the correct prediction of TE noise for rotating blades.

%The model assumes that the kernel can be exactly
%decomposed so that a fully separably solution can be obtained, which appears to
%be not true as will be shown later. 

A classical way to incorporate the 3D effects is to use the two-step approach
used in Amiet's model~\citep{Amiet1976b}, where the surface pressure due to the
gust scattering by a serrated semi-infinite plate is calculated first, and the
far-field sound is calculated subsequently using a surface integral assuming a
finite plate. To do that, it is crucial to obtain the near-field scattered
pressure on the plate surface. This poses a great difficulty as the powerful
method of the steepest descent cannot be used to evaluate the inverse Fourier
transform as used in \citet{Ayton2018d}. Considering acoustic reciprocity, this
is in fact equivalent to calculating the Green's function for the acoustic
scattering by serrated edges, where the acoustic source, instead of the
observer, is placed in the near field. Obtaining such a Green's function would
enable a TE noise model to be developed that is both three-dimensionally
accurate and computationally efficient. Moreover, the Green's function itself
is fundamentally important in many important aspects concerning TE noise.
First, this would open the possibility of examining the consequences of many
assumptions that have been open to heated debate, such as the validity of
frozen turbulence that has been called into question in a number of recent
studies~\citep{Ragni2019,Zhou2020}. Second, the Green’s function would provide
a more intuitive understanding of the effects of serrations by showing the
scattering characteristics of simple sound sources, thereby lending insights
into the physical mechanism of noise reduction by using serrations, and more
importantly informing new techniques of suppressing TE noise. Last but not the
least, the Green's function would permit a direct comparison between analytical
scattering models and experiments. The point-source induced sound can be
readily measured in the laboratory using laser-induced monopoles. Although TE
noise modelling has improved significantly, it is yet to see robust agreement
between trailing-edge noise models and experiments. This is difficult,
especially when realistic aerofoil geometries are considered. Often this is
because the surface pressure statistics needed in the noise prediction model
are rather difficult to be obtained accurately. With a controlled simple
acoustic source, we can readily assess whether any deviations that exist
between models and experiments are introduced by the scattering model and its
underlying assumptions or by the turbulent pressure fluctuations statistics.

%It is not clear whether such a discrepancy is caused by the deficiencies of
%the scattering model or the inaccurate modelling of the turbulence in the
%boundary layer, because we cannot compare the sound induced by a
%single-component gust with the corresponding model as such a gust is hard to
%be generated in the laboratory. On the other hand, acoustic measurement using
%a point source, i.e. the Green’s function for the scattering problem, can be
%carried out readily in the experiment using laser induced monopoles and
%therefore can be used to validate the modelling techniques. 

Although important, such an analytical Green's function remains unknown. As
mentioned above, a tailored Green's function was proposed by Howe in
1991~\citep{Howe1991a}, but it has been shown problematic, especially when the
serration is sharp. In this paper, we aim to develop such a Green's function
analytically by using the Wiener-Hopf method. \add{Due to mathematical
symmetry, with proper transformations it would also be applicable to the
scattering by serrated leading edges~\citep{Amiet1975,Amiet1976b,Lyu2017c}.}
This paper is structured as follows. Section~\ref{sec:GreensFunction}
introduces the simplified model and develops the Green's function.
Section~\ref{sec:Validation} validates the Green's function by performing
numerical integrations and Finite Element Method (FEM) computations using
COMSOL. In section~\ref{sec:Results} we show the noise directivity for a point
source located near the serrated trailing edge of the flat plate due to the use
of serrations and examine the effects of varying the frequency, serration
amplitude, source position and Mach number, respectively. The final section
concludes the paper and lists some future work.

\section{Analytical derivation}
\label{sec:GreensFunction}
\begin{figure}
    \centering
    \includegraphics[width=0.9\linewidth]{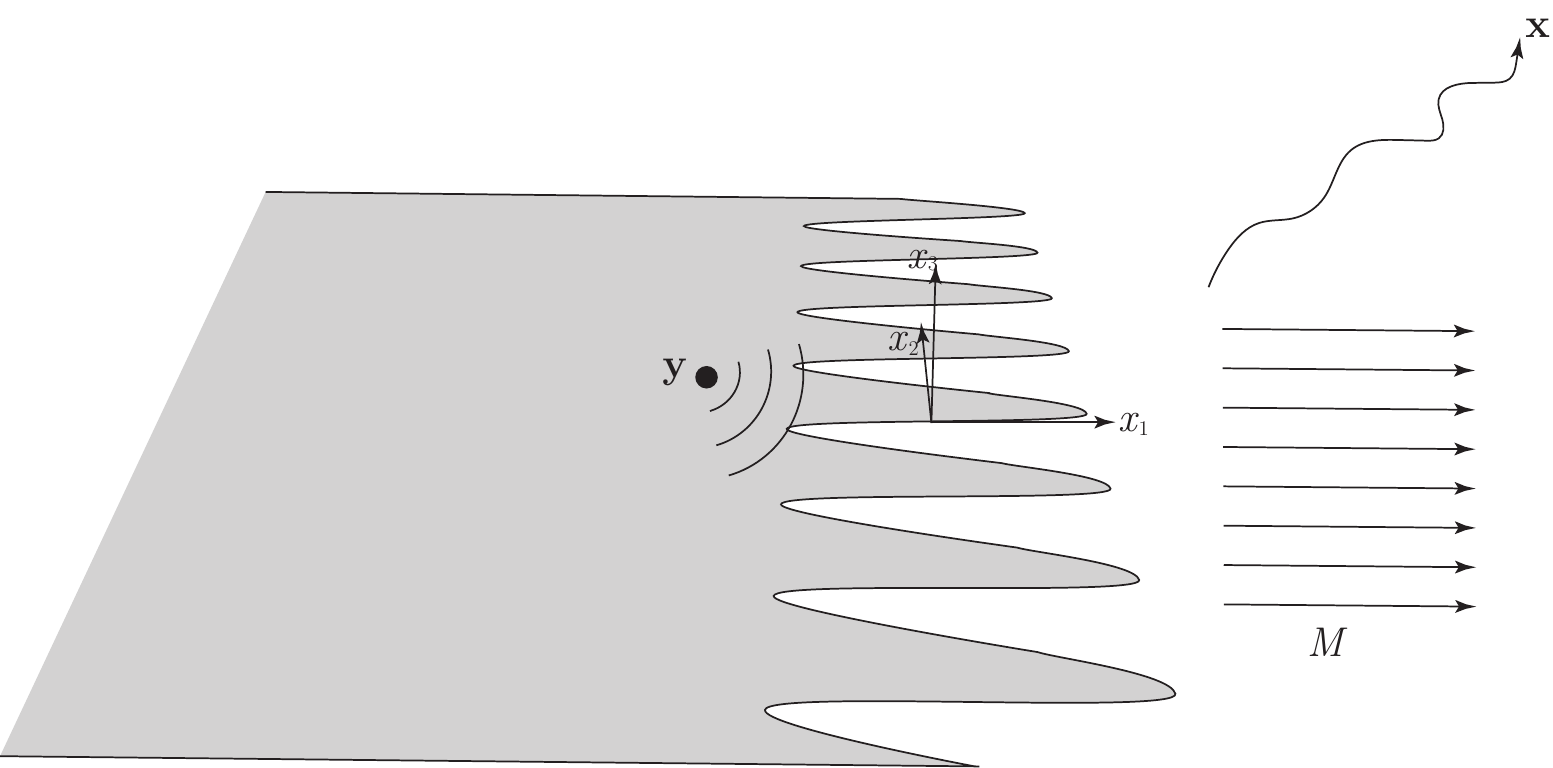}
    \caption{Schematic illustration of the Green's function problem. The source
	represented by the black dot in the diagram is located near the edge at
	$\boldsymbol{y}$, and the observer is located in the far field at
    $\boldsymbol{x}$.}
    \label{fig:GreenFunctionSerratedPlate}
\end{figure}

To allow analytical progression, we start from a simplified model that is
widely used in the literature~\citep{Howe1978,Lyu2016a}, i.e. the aerofoil is
simplified as a flat plate placed in a uniform flow aligned in the streamwise
direction, as shown in figure~\ref{fig:GreenFunctionSerratedPlate}. As
mentioned in section~\ref{sec:Introduction}, the Green's function would be
applicable to both trailing-edge and leading-edge scattering because of
mathematical symmetry. In this paper, we use the trailing-edge scattering as an
example. The flat plate is assumed to be semi-infinite, i.e. the leading edge
\add{extends to} the upstream infinity and both side edges are also infinitely
far away, so only the serrated trailing edge needs to be considered.  We
restrict our analysis to periodic serrations with a wavelength of
$\tilde{\lambda}$. The problem is non-dimensionalized using the serration
wavelength $\tilde{\lambda}$, the speed of sound $\tilde{c}$ and the fluid
density $\tilde{\rho}$. Note we have used the symbols with a tilde to denote
dimensional variables, whereas those without represent non-dimensional
variables. We will adhere to this convention throughout this paper unless
explicitly noted otherwise. In terms of the non-dimensional variables, the
serration has a wavelength $1$ and \add{half root-to-tip} amplitude $h$, and
the uniform flow from left to right has a dimensionless velocity $M$, which is
just the Mach number.

A Cartesian coordinate system shown in
figure~\ref{fig:GreenFunctionSerratedPlate} is used in the analysis, where
$x_1$, $x_2$, and $x_3$ denote the dimensionless streamwise, spanwise, and
normal-to-plate coordinates, respectively. In such a coordinate system, the
profile of the serration, or the trailing edge of the plate, can be described
by the periodic function $x_1 = hF(x_2)$, where $F(x_2)$ obtains a maximum
value of $1$ and a minimum value of $-1$, respectively. Under the harmonic
assumption of $\e^{-\i \omega t}$, where $\omega$ is the non-dimensionalized
angular frequency, the Green's function $G(\boldsymbol{x};
\boldsymbol{y},\omega)$ satisfies the following inhomogeneous convective
equation~\citep{Amiet1976a,Lyu2016a}
\begin{equation}
    \left(\beta^2 \frac{\partial^2 }{\partial x_1^2} 
	+ \frac{\partial^2 }{\partial x_2^2} 
	+ \frac{\partial^2 }{\partial x_3^2} 
	+ 2\i k M \frac{\partial }{\partial x_1} 
    + k^2 \right) G(\boldsymbol{x}; \boldsymbol{y}, \omega) =
    \delta(\boldsymbol{x}-\boldsymbol{y}),
    \label{equ:ConvectiveWaveEquation}
\end{equation}
and the boundary condition 
\begin{equation}
    \left.\frac{\partial G}{\partial x_3}\right|_{x_3 =\pm 0} = 0, \quad x_1 <
	hF(x_2)
\end{equation}
where, $\beta = \sqrt{1 - M^2}$ and $k = \omega/c$, and as shown in
figure~\ref{fig:GreenFunctionSerratedPlate}, $\boldsymbol{y}$ denotes the
source position, while $\boldsymbol{x}$ denotes the observer position. 

Note the observer location $\boldsymbol{x}$ is often in the far field,
therefore a standard technique is to use the reciprocal theorem to calculate
the adjoint Green's function $G^a(\boldsymbol{y}; \boldsymbol{x}; \omega)\equiv
G(\boldsymbol{x}; \boldsymbol{y}, \omega)$ so that the advantage of a plane
wave incidence can be taken. However, as we aim to include the mean-flow
convection effect in this paper, i.e. $M \ne 0$,
(\ref{equ:ConvectiveWaveEquation}) is no longer self-adjoint. In other
words, the adjoint Green's function $G^a(\boldsymbol{y}; \boldsymbol{x},
\omega)$ does not satisfy (\ref{equ:ConvectiveWaveEquation}).
Nevertheless, it can be shown that the equation that $G^a(\boldsymbol{y};
\boldsymbol{x}, \omega)$ does satisfy differs from
(\ref{equ:ConvectiveWaveEquation}) only by the sign in front of the term
$2\i kM \frac{\partial }{\partial x_1}$, i.e. 
\begin{equation}
    \left(\beta^2 \frac{\partial^2 }{\partial y_1^2} 
	+ \frac{\partial^2 }{\partial y_2^2} 
	+ \frac{\partial^2 }{\partial y_3^2} 
	- 2\i k M \frac{\partial }{\partial y_1} 
    + k^2 \right) G^a(\boldsymbol{y}; \boldsymbol{x}, \omega) =
    \delta(\boldsymbol{y}-\boldsymbol{x}).
    \label{equ:AdjointWaveEquation}
\end{equation}
Physically, this is equivalent to solving the acoustic pressure at
$\boldsymbol{y}$ while the point source is at $\boldsymbol{x}$, assuming a
uniform flow of Mach number $M$ travels from right to left. In other words, the
problem can be cast as ``reciprocal" by reversing the uniform mean flow. 

Because $\boldsymbol{x}$ is in the far field, the incidence wave from the
source $\boldsymbol{x}$ can be approximated by a plane wave, whose amplitude
depends on the distance between $\boldsymbol{x}$ and $\boldsymbol{y}$. Because
of linearity, we can start with an incident wave of magnitude $1$, i.e.
\begin{equation}
    p_{in} = \e^{-\i k_1 y_1 / \beta} 
    \e^{\i\frac{kM}{\beta^2}y_1} 
    \e^{-\i (k_2 y_2 + k_3 y_3)}, 
    \label{equ:incidentWave}
\end{equation}
where $k_1$ and $k_2$ are constants related to the radiation angle, the precise
definition of which will be given later, and $k_3 = \sqrt{(k/\beta)^2 - k_1^2 -
k_2^2}$. It can be verified that (\ref{equ:incidentWave}) satisfies the
homogeneous version of (\ref{equ:AdjointWaveEquation}). We decompose the
total adjoint pressure field $G^a = p_{in} + p_r + R_s$, where the
hypothetically reflected wave $p_r$ off an infinite flat plate is defined as
$p_r = p_{in}(y_1, y_2, -y_3)$ and $R_s$ is the reflection-removed scattered
pressure field. We could also have decomposed the pressure field as $G^a =
p_{in} +G_s$, and this approach is shown in Appendix B. Note however no matter
which decomposition is used, it should in no way affect the final solution.

The reflection-removed scattered wave
$R_s$ satisfies
\begin{equation}
    \beta^2 \frac{\partial^2 R_s}{\partial y_1^2} 
    + \frac{\partial^2 R_s}{\partial y_2^2} 
    + \frac{\partial^2 R_s}{\partial y_3^2} 
    - 2 \i k M \frac{\partial R_s}{\partial y_1} 
    + k^2 R_s = 0,
    \label{equ:WaveEquationMainText}
\end{equation}
and the following boundary conditions due to the periodicity of the serrations ~\citep{Ayton2018d}, 
\begin{IEEEeqnarray}{rCl}
    \left.\frac{\partial R_s}{\partial y_3}\right|_{y_3=0} 
	   & =&  0,  \quad y_1 < hF(y_2);\IEEEyesnumber\IEEEyessubnumber\\
	   \left.R_s\right|_{y_3 = 0} 
	   &=& -\e^{-\i k_1 y_1 / \beta}
	   \e^{\i\frac{kM}{\beta^2}y_1} 
	   \e^{-\i k_2 y_2}, \quad y_1 > hF(y_2); \IEEEyessubnumber\\
	   \left.R_s\right|_{y_2 = 0} 
	   &=& \left.R_s\right|_{y_2 = 1} \e^{\i k_2}; \IEEEyessubnumber\\
	       \left.\frac{\partial R_s}{\partial y_2}\right|_{y_2 = 0} 
	   &=& \left.\frac{\partial R_s}{\partial y_2}\right|_{y_2 = 1}
	       \e^{\i k_2}. \IEEEyessubnumber
\end{IEEEeqnarray}

Eliminating the first-order term in (\ref{equ:WaveEquationMainText}) by the
transformation $R_s = \bar{R}_s \e^{\i k M y_1 / \beta^2}$, we obtain
\begin{equation}
    \beta^2 \frac{\partial^2 \bar{R}_s}{\partial y_1^2} 
    + \frac{\partial^2 \bar{R}_s}{\partial y_2^2} 
    + \frac{\partial^2 \bar{R}_s}{\partial y_3^2} 
    + \left(\frac{k}{\beta}\right)^2 \bar{R}_s = 0.
    \label{equ:StretchedWaveEquationMainText}
\end{equation}
Earlier work~\citep{Ayton2018d} often uses the non-orthogonal coordinate
transformation $\bxi_1 = (y_1 - h F(y_2)) / \beta$, $\bxi_2 = y_2$ and $\bxi_3 =
y_3$ to enable the use of separation of variables. We show that this coordinate
transformation is not necessary and the same Wiener-Hopf equation can be
obtained by using the Fourier transform directly. We follow this approach here.
Introducing the stretched coordinate $\bxi_1 = y_1 / \beta$, $\bxi_2= y_2$,
$\bxi_3 = y_3$, we see that the governing equation reduces to
    \begin{equation}
	\frac{\partial^2 \bar{R}_s}{\partial \bxi_1^2} + \frac{\partial^2
	\bar{R}_s}{\partial \bxi_2^2} + \frac{\partial^2 \bar{R}_s}{\partial
    \bxi_3^2} + \bar{k}^2 \bar{R}_s = 0,
    \label{equ:HelmholtzEquation}
\end{equation}
where the stretched constants are defined as $\bar{k} = k/\beta$. Now the
boundary conditions read
\begin{IEEEeqnarray}{rCl}
    \label{BCAfterChangeVariableMainText}
    \frac{\partial \bar{R}_s}{\partial \bxi_3}|_{\bxi_3=0} 
	&=& 0, \quad \bxi_1 < \bar{h} F(\bxi_2);\IEEEyesnumber\IEEEyessubnumber\\
	\bar{R}_s|_{\bxi_3 = 0} 
	&=& -\e^{-\i (k_1 \bxi_1 + k_2 \bxi_2)},
	\quad \bxi_1 > \bar{h} F(\bxi_2); \IEEEyessubnumber\\
	\bar{R}_s|_{\bxi_2 = 0} 
	&=& \bar{R}_s|_{\bxi_2 = 1} \e^{\i k_2}; \IEEEyessubnumber  \\
	\frac{\partial \bar{R}_s}{\partial \bxi_2}|_{\bxi_2 = 0} 
	&=& \frac{\partial \bar{R}_s}{\partial \bxi_2}|_{\bxi_2 = 1}\e^{\i k_2}, 
	\IEEEyessubnumber
\end{IEEEeqnarray}
where $\bar{h}$ is defined as $\bar{h} = h/\beta$. We can now perform the
Fourier transform along the $\bxi_1$ direction, i.e.
\begin{equation}
    \mathcal{R}(s, \bxi_2, \bxi_3) = \int_{-\infty}^{\infty} \bar{R}_s(\bxi_1, \bxi_2,
    \bxi_3) \e^{\i s \bxi_1} \ud \bxi_1.
\end{equation}
Function $\mathcal{R}(s,
\bxi_2, \bxi_3)$ can be decomposed into two parts, i.e. 
\begin{IEEEeqnarray}{rCl}
    &&\mathcal{R}(s, \bxi_2, \bxi_3) 
    =
    \int_{-\infty}^{\bar{h}F(\bxi_2)}\bar{R}_s(\bxi_1,
    \bxi_2, \bxi_3) \e^{\i s \bxi_1} \ud \bxi_1
    + \int_{\bar{h}F(\bxi_2)}^\infty \bar{R}_s(\bxi_1,
    \bxi_2, \bxi_3) \e^{\i s \bxi_1} \ud \bxi_1\nonumber\\
    &&= 
    \int_{-\infty}^0\bar{R}_s(\bxi_1+\bar{h}F(\bxi_2),
    \bxi_2, \bxi_3) \e^{\i s (\bxi_1+\bar{h}F(\bxi_2))} \ud \bxi_1
    + \int_{0}^\infty \bar{R}_s(\bxi_1+\bar{h}F(\bxi_2),
    \bxi_2, \bxi_3) \e^{\i s (\bxi_1+\bar{h}F(\bxi_2))} \ud
    \bxi_1\nonumber\\
    &&= 
    \e^{\i s \bar{h} F(\bxi_2)} \left( \mathcal{R}^-(s, \bxi_2, \bxi_3) +
    \mathcal{R}^+(s, \bxi_2, \bxi_3) \right) 
\end{IEEEeqnarray}
where functions $\mathcal{R}^-$ and $\mathcal{R}^+$ are complex functions that
are analytical in the lower and upper half $s$ planes, respectively. A similar
Fourier transform (and decomposition) is applied to the function $\partial
\bar{R}_s / \partial \bxi_3$, the result of which will be denoted by
$\mathcal{R}^\prime$ in the rest of the paper. 

Equation~(\ref{equ:HelmholtzEquation}) then reduces to 
\begin{equation}
    \frac{\partial^2 \mathcal{R}}{\partial \bxi_2^2}%
    + \frac{\partial^2 \mathcal{R}}{\partial \bxi_3^2}%
    + (\bar{k}^2-s^2) \mathcal{R} = 0.
    \label{equ:AfterFourier}
\end{equation}
Equation~(\ref{equ:AfterFourier}) is the standard Helmholtz equation and its
solution can be found through the usual method of separation of variables.
\add{After using the last two boundary conditions shown in
(\ref{BCAfterChangeVariableMainText}), we show that (\ref{equ:AfterFourier})
can be solved for $\bxi_3>0$ (the corresponding result for $\bxi_3<0$ is
similar due to antisymmetry) to yield}
\begin{equation}
    \mathcal{R}(s, \bxi_2, \bxi_3) = \sum_{n = -\infty}^{\infty} A_n(s)
    \e^{-\gamma_n \bxi_3} \e^{\i \chi_n \bxi_2},
    \label{equ:determinedFormMainText}
\end{equation}
where $\chi_n = 2n\pi - k_2$, $\gamma_n = \sqrt{s^2 - \kappa_n^2}$, and
$\kappa_n = \sqrt{\bar{k}^2 - \chi_n^2}$. We see that $\kappa_n$ denotes the
wavenumber in the $\bxi_1-\bxi_3$ plane and when $n=0$ it is equal to $\sqrt{k_1^2
+ k_3^2}$ . The complex function $A_n(s)$ will need to be determined by making
use of the first two boundary conditions shown in
(\ref{BCAfterChangeVariableMainText}) by using the Wiener-Hopf
method, i.e.
\begin{IEEEeqnarray}{rCl}
   \label{equ:RsBC}
   \mathcal{R}^\prime(s, \bxi_2,  0) 
   &=& \e^{\i s \bar{h} F(\bxi_2) }\sum_{n=-\infty}^{\infty}
   \mathcal{R}_n^{\prime+}(s)\e^{-i s \bar{h} F(\bxi_2)} \e^{\i \chi_n \bxi_2};
   \IEEEyesnumber\IEEEyessubnumber\\
   \mathcal{R}(s,\bxi_2, 0) 
   &=& \e^{\i s \bar{h} F(\bxi_2) } \left(
       \sum_{n=-\infty}^{\infty}
       \mathcal{R}_{n}^-(s) \e^{-i s \bar{h} F(\bxi_2)} \e^{\i \chi_n \bxi_2}
       - \frac{\i}{s - k_1} \e^{-\i (k_1 \bar{h} F(\bxi_2)+ k_2 \bxi_2)}
   \right),\IEEEeqnarraynumspace\IEEEyessubnumber
   \label{equ:RsBCb}
\end{IEEEeqnarray}   
where $\mathcal{R}_n^{\prime +}(s)$ and $\mathcal{R}_n^-(s)$ are the expansion
coefficients of functions $\mathcal{R}^{\prime +}(s, \bxi_2, 0)$ and
$\mathcal{R}^-(s, \bxi_2, 0)$ using the basis functions $\e^{-\i s \bar{h}
F(\bxi_2)} \e^{\i \chi_n \bxi_2}, n =0, \pm 1, \pm 2 ...$, and they are unknown at
this stage. The last exponential term in the parenthesis of
(\ref{equ:RsBCb}) can also be expanded and the resulting coefficients
are denoted by $E_n(s)$. $E_n(s)$ can be found to be 
\begin{equation}
    E_n(s) = \int_0^1 \e^{\i (s-k)\bar{h}F(\bxi_2)}\e^{-\i 2 n \pi \bxi_2} \ud
    \bxi_2.
    \label{equ:EnDefinition}
\end{equation}
Note that $E_n(s)$ can be arbitrary because no restriction on $F(\bxi_2)$ has
been imposed apart from it being periodic. For any arbitrary piecewise
linear functions, $E_n(s)$ can be integrated analytically. For example, for the
conventional sawtooth serration profile defined by (in one period)
\begin{equation}
    F(\bxi_2)=    
    \begin{cases}
	4 \bxi_2, \quad & -\frac{1}{4} < \bxi_2 < \frac{1}{4} \\
	-4 \bxi_2 + 2, \quad & \frac{1}{4} < \bxi_2 < \frac{3}{4},
    \end{cases}
\end{equation}
$E_n(s)$ can be found as
\begin{equation}
    E_n(s)= \frac{4 (s - k_1)\overline{h} 
    \sin\left((s-k_1)\overline{h} - n\pi/2\right)}
    {4 (s-k_1)^2\overline{h}^2 - n^2 \pi^2}.
    \label{equ:EnSawtooth}
\end{equation}

Upon comparing (\ref{equ:determinedFormMainText}) and (\ref{equ:RsBC}) and
making use of orthogonality of the basis functions $\e^{-\i s \bar{h} F(\bxi_2)}
\e^{\i \chi_n \bxi_2}, n =0, \pm 1, \pm 2 ...$, we arrive at the following
matching conditions for mode $n$, i.e.
\begin{IEEEeqnarray}{rCl}
 -\gamma_n A_n(s)
&=&  \mathcal{R}^{\prime + }_n(s), \IEEEyesnumber\IEEEyessubnumber \\
A_n(s) 
&=& \mathcal{R}_{n}^-(s) - \frac{\i}{s -k_1} E_n(s).\IEEEyessubnumber
\end{IEEEeqnarray}
We can proceed by eliminating $A(s)$ and arrive at the Wiener-Hopf equation
\begin{equation}
    \gamma_n \left(\mathcal{R}^-_n(s) - \frac{\i}{s-k_1} E_n(s)\right) + 
    \mathcal{R}_n^{\prime +}(s) = 0.
    \label{equ:WienerHopfEquationMainText}
\end{equation}
The function $E_n(s)$ causes much difficulty in the kernel decomposition.
Recent approach~\citep{Ayton2018d} assumes that both $\mathcal{R}_n^-(s)$ and
$\mathcal{R}_n^{\prime +}(s)$ contain the factor $E_n(s)$ so that a kernel
factorization can proceed. However, we find that this assumption appears not
true, in particular this leads to results that do not strictly satisfy the
boundary conditions. Moreover, as mentioned above, the results obtained by
decomposing the total pressure field as either $G^a = p_{in} + G_s$ or $G^a =
p_{in} + p_r + R_s$ should yield no difference to the final solution. However,
it can be verified that if the assumption that $E_n(s)$ is a factor in
$R_n^-(s)$ and $R_n^{\prime+}(s)$ is used, the two methods would yield
different solutions (the two are only equal to each other for mode $n=0$, see
Appendix~\ref{app:AnotherDecomposition} for details), which signals a potential
problem with the underlying assumption. In fact, from
(\ref{BCAfterChangeVariableMainText}) and (\ref{equ:RsBC}) we see that $E_n(s)$
represents the variation of the incident pressure on the edge. As
$\mathcal{R}_n^-(s)$ denotes the scattered pressure upstream of the trailing
edge, if $\mathcal{R}_n^-(s)$ had the same $E_n(s)$ factor as the incident
wave, the scattering problem would need to be homogeneous in the spanwise
direction. This can only be guaranteed if the trailing edge is a straight (or
swept) edge. For serrated edges, the homogeneity condition is not satisfied and
the $E_n(s)$ variation in $\mathcal{R}_n^-(s)$ cannot be guaranteed. 

However, as the frequency increases, the acoustic wavelength becomes
increasingly short, and the scattered pressure variation on the edge is
expected to become increasingly localized and dominated by the incident phase
variation; the assumption of $E_n(s)$ dependence may be approximately valid in
the high frequency limit. \add{Serrations are known to be more effective as
frequency increases (see for example \citet{Howe1991a}, \citet{Gruber2012} and
\citet{Lyu2016a}), and more importantly, it is the hydrodynamic wavelength that
characterises the incoming (gust) length scale in TE noise modelling,} the
localized scattering is more likely to be valid. Therefore, in the following
part of this paper we focus on this high frequency regime aiming to develop a
closed-form analytical Green's function, which can be used to develop a
three-dimensional TE noise model. 
%and examine to what extent this approximates the exact solution. 

The kernel is the standard $\gamma_n = \sqrt{s^2 - \kappa_n^2}$, and once
$E_n(s)$ is removed from both $\mathcal{R}_n^-(s)$ and $\mathcal{R}_n^{\prime
+}(s)$ \add{it becomes a routine procedure to be factorised as} $\sqrt{s -
\kappa_n}\sqrt{s + \kappa_n}$. $A_n(s)$ can then be approximated by

\begin{figure}
    \centering
    \includegraphics[width=0.8\textwidth]{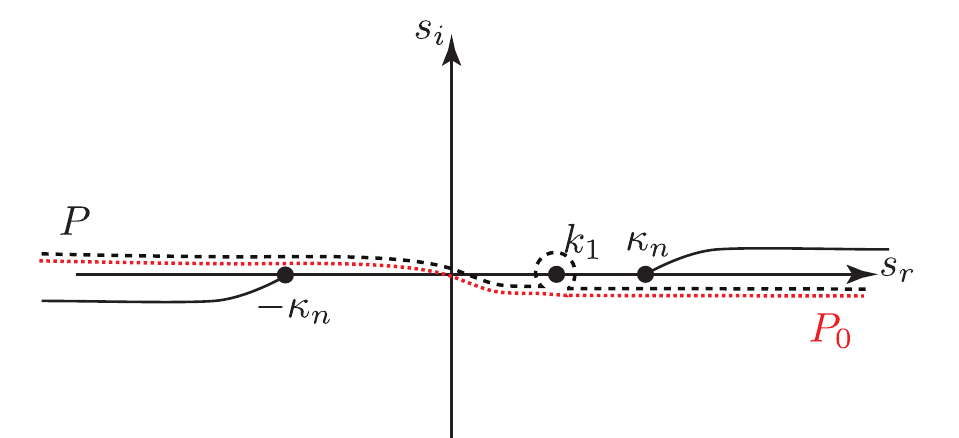}
    \caption{The integral path $P$ in (\ref{equ:finalIntegralEquation}),
    which passes around a simple pole at $s=k_1$ and two branch points at $s
    =\pm\kappa_n$. Note the branch point $\kappa_n$ can be an imaginary number
    depending on the value of $n$, which however does not affect the
    analyticity of the integrand along the integral path. The integral along
    path $P$ is equivalent to that along $P_0$ minus a residue contribution
    around $s=k_1$.}
    \label{fig:integralPath}
\end{figure}

%Upon dividing both sides of
%(\ref{equ:WienerHopfEquation}) by $\sqrt{s+\kappa_n}$ and cancelling out
%the pole in the second term of (\ref{equ:WienerHopfEquation}), we have 
%\begin{equation}
%    \begin{aligned}
%    \sqrt{s - \kappa_n} \mathcal{R}^-_n(s) 
%    + \frac{k_3 E_n(s)}{(s-k_1)}  
%    \frac{1}{\sqrt{k_1 + \kappa_n}}
%    & =
%    - \frac{\mathcal{R}_n^\prime(s)}{\sqrt{s + \kappa_n}} 
%    - \frac{k_3 E_n(s)}{(s-k_1)}  
%    \left(\frac{1}{\sqrt{s + \kappa_n}} - \frac{1}{\sqrt{k_1 +
%    \kappa_n}}\right) \\
%    & \equiv E(s).
%    \end{aligned}
%    \label{equ:Match}
%\end{equation}
%It is easy to verify that $E(s) \equiv 0$ by invoking the Louville Theorem,
%and we are left with, 
\begin{equation}
    A(s) = - \frac{1}{\gamma_n} \mathcal{R}^{\prime+}_n(s) 
    = -\frac{\i}{\sqrt{s - \kappa_n}}
    \frac{\sqrt{k_1 - \kappa_n}}{s-k_1} E_n(s)  
    \label{equ:A}
\end{equation}
Substituting (\ref{equ:A}) into (\ref{equ:determinedFormMainText}) and
taking the inverse Fourier transform yields
\begin{equation}
    \bar{R}_s(\bxi_1, \bxi_2, \bxi_3)) 
    =  
    \sum_{n = -\infty}^{\infty}
    -\i(\sqrt{k_1 - \kappa_n})
    \e^{\i \chi_n \bxi_2}
    \frac{1}{2\pi}
    \int_{-\infty}^{\infty}
    \frac{E_n(s)}{s-k_1}  
    \frac{1}{\sqrt{s - \kappa_n}}
    \e^{-\i s \bxi_1  - \gamma_n \bxi_3} 
    \ud s. 
\end{equation}
Let $r =  \sqrt{(y_1/\beta)^2 + y_3^2}$, and $\cos\theta = y_1 / (\beta r)$, we
have finally 
\begin{IEEEeqnarray}{rCl}
    R_s(r,\theta, y_2) = \frac{1}{2\pi}\e^{\i k M y_1 / \beta^2}
    \sum_{n = -\infty}^{\infty} &&
    -\i(\sqrt{k_1 - \kappa_n})
    \e^{\i \chi_n y_2} 
    \nonumber\\
		       && \times \left[
			   \int_{-\infty}^{\infty}
			   \frac{E_n(s)}{s-k_1}  
			   \frac{1}{\sqrt{s - \kappa_n}}
			   \e^{(-i s  \cos \theta  - \gamma_n  \sin\theta) r} \ud s  
		       \right], \IEEEeqnarraynumspace
		       \label{equ:finalIntegralEquation}
\end{IEEEeqnarray}
where $E_n(s)$ is given by (\ref{equ:EnDefinition}), and the integral is along
the path $P$ shown in figure~\ref{fig:integralPath}. Note that the integrand in
(\ref{equ:finalIntegralEquation}) has a pole at $s=k_1$ and two branch points
at $s=\pm\kappa_n$. The integral path $P$ has to pass above the pole at $s=k_1$
due to the analyticity requirement. It is, however, equivalent to integrating
(\ref{equ:finalIntegralEquation}) along the path $P_0$ shown in
figure~\ref{fig:integralPath}, provided that the residue contribution from the
pole is subtracted. We see from (\ref{equ:EnSawtooth}) that $E_n(k_1) =
\delta_{n_1}$, \add{therefore it is convenient to calculate the residue,} which
is precisely the hypothetical reflected wave off an infinite flat plate
$p_{r}$. Consequently, the total scattered field $G_s$ can be directly
calculated by integrating (\ref{equ:finalIntegralEquation}) along the path
$P_0$ instead, i.e.
\begin{IEEEeqnarray}{rCl}
    G_s(r,\theta, y_2) = \frac{1}{2\pi}\e^{\i k M y_1 / \beta^2}
    \sum_{n = -\infty}^{\infty} &&
    -\i(\sqrt{k_1 - \kappa_n})
    \e^{\i \chi_n y_2} 
    \nonumber\\
		       && \times \left[
			   \int_{-\infty}^{\infty}
			   \frac{E_n(s)}{s-k_1}  
			   \frac{1}{\sqrt{s - \kappa_n}}
			   \e^{(-i s  \cos \theta  - \gamma_n  \sin\theta) r} \ud s  
		       \right], \IEEEeqnarraynumspace
		       \label{equ:GsIntegral}
\end{IEEEeqnarray}
where the integral path in (\ref{equ:GsIntegral}) is given by $P_0$ as
shown in figure~\ref{fig:integralPath}.

To obtain a closed-form analytical Green's function, the integral in
(\ref{equ:GsIntegral}) has to be evaluated analytically. Note that $E_n(s)$ is
arbitrary, but for all piecewise linear serration profiles $E_n(s)$ can be
evaluated analytically. If the far-field scattered pressure is of interest,
i.e. $r\to\infty$, (\ref{equ:GsIntegral}) can be quickly evaluated
asymptotically by the powerful method of the steepest descent, as shown by
\citet{Ayton2018d}. However, as we seek the Green's function, it is the
near-field scattered pressure that is of our interest. The steepest descent
method can no longer be used, and the contour integral in
(\ref{equ:GsIntegral}) must be integrated exactly. We show that for all
piecewise linear profiles, the above complex contour integral can be integrated
exactly to yield closed-form analytical solutions. We use the conventional
sawtooth serration as an example in the rest of the paper, whereas the Green's
functions for other common piecewise linear functions are given in
Appendix~\ref{app:OtherGs}.

\begin{figure}
    \centering
    \includegraphics[width=0.85\textwidth]{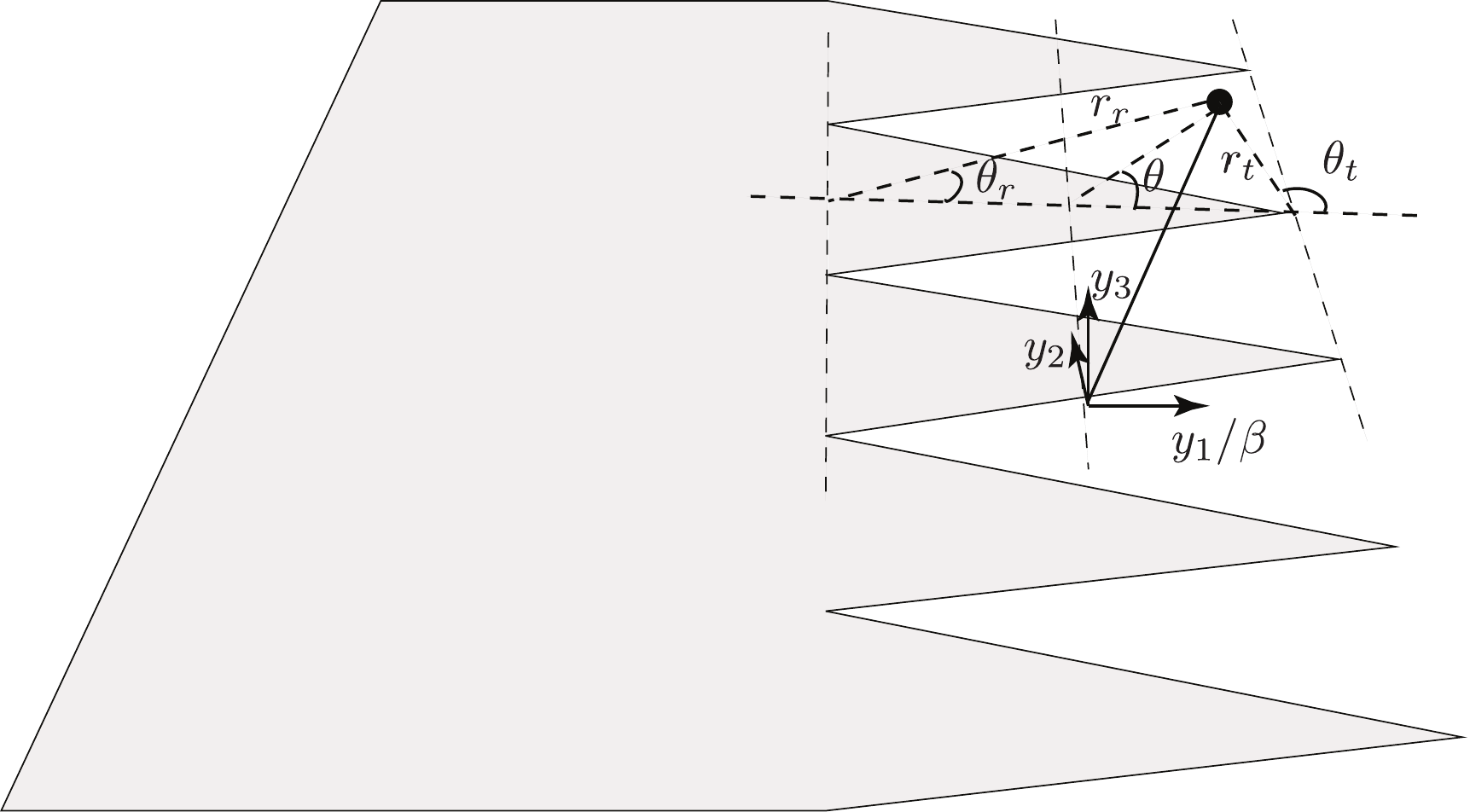}
    \caption{The definition of the two geometrical angles $\theta_t$ and
    $\theta_r$ and their corresponding radial coordinates $r_t$ and $r_i$.}
    \label{fig:GeoAngleDef}
\end{figure}

\add{We begin by noting that for conventional sawtooth serrations, $E_n(s)$ are
given by (\ref{equ:EnSawtooth}), in which the sine functions can be
expanded using exponential functions.} To facilitate a compact notation, we
define two auxiliary local polar coordinate frames, i.e. $(r_t, \theta_t)$ and
$(r_r, \theta_r)$ in the stretched $y_1 / \beta - y_3$ plane (i.e.
$\bxi_1-\bxi_3$ plane), as shown in figure~\ref{fig:GeoAngleDef}. Here the
stretch factor $\beta$ is to account for the background uniform flow, and when
$M=0$ the stretched plane is just the physical $y_1-y_3$ plane. We see that
$\theta_t$ and $\theta_r$ represent the geometric angles of the observer with
respect to the tip and root of the serration in the stretched $y_1/\beta-y_3$
plane, respectively, while $r_t$ and $r_r$ represent their corresponding radial
coordinates, respectively. \add{With these definitions, we can obtain}
\begin{IEEEeqnarray}{rCl}
    r_t &=& \sqrt{r^2 + \bar{h}^2 - 2 r \bar{h} \cos\theta},
    \IEEEyesnumber\IEEEyessubnumber\\
    \theta_t &=& \arccos \left[(r\cos\theta -
    \bar{h})/r_t\right].\IEEEyessubnumber
\end{IEEEeqnarray}
And similarly we have
\begin{IEEEeqnarray}{rCl}
    r_r &=& \sqrt{r^2 + \bar{h}^2 + 2 r \bar{h} \cos\theta}, \nonumber\\
    \theta_r &=& \arccos \left[(r\cos\theta + \bar{h})/r_r\right].\nonumber
\end{IEEEeqnarray}

\add{Expanding the sine functions in (\ref{equ:EnSawtooth}) into
exponential functions, we can show that }
\begin{IEEEeqnarray}{rCl}
    G_s(r, \theta, y) &=& \frac{1}{2\pi}\e^{\i k M y_1 / \beta^2}\sum_{n = -\infty}^{\infty}
    -\i(\sqrt{k_1 - \kappa_n})
    \e^{\i \chi_n y_2}\nonumber\\
		      && \times\left( \exp\left( -\i(k_1
			  \bar{h}+\frac{n\pi}{2})\right) H_n(r_t, \theta_t) 
		      - \exp\left( \i(k_1
		  \bar{h}+\frac{n\pi}{2})\right)H_n(r_r, \theta_r)\right),
		  \IEEEeqnarraynumspace 
		  \label{equ:Gs}
\end{IEEEeqnarray}
where
\begin{equation}
	H_n(r_i, \theta_i) = \int_{-\infty}^{\infty}
	\frac{-2\i\overline{h}}{(2(s-k_1)\overline{h})^2-(n\pi)^2}
	\frac{1}{\sqrt{s - \kappa_n}}
	\e^{(-i s \cos \theta_i  - \gamma_n  \sin\theta_i)r_i} \ud s 
	\label{equ:Hn}
\end{equation}
and $(r_i, \theta_i)$ can take the value of either $(r_t, \theta_t)$ or $(r_r,
\theta_r)$.

To integrate (\ref{equ:Hn}), when $n\ne 0$ we may expand the first
factor of the integrand as partial fractions, i.e. 
\begin{equation}
	\frac{-2\i\overline{h}}{(2(s-k_1)\overline{h})^2-(n\pi)^2} 
	=-
	\frac{\i \overline{h}}{n \pi}
	\left[\frac{1}{2(s-k_1)\overline{h} - n\pi} -
	\frac{1}{2(s-k_1)\overline{h} + n\pi}
    \right].
\end{equation}
Equation~(\ref{equ:Hn}) can then be written as the difference between two
integrals, i.e. 
\begin{equation}
    H_n(r_i, \theta_i) = -\frac{\i}{2n\pi}\big[D_n^+(r_k, \theta_i) -
	D_n^-(r_k, \theta_i)\big]
\end{equation}
where
\begin{equation}
    D_n^\pm(r_k, \theta_i) = 
    \int_{-\infty}^{\infty}
    \frac{1}{s-(k_1 \pm n\pi/2 \overline{h})}
    \frac{1}{\sqrt{s - \kappa_n}}
    \e^{(-i s \cos \theta_i  - \gamma_n  \sin\theta_i)r_i} \ud s.
    \label{equ:FinalFraction}
\end{equation}

%In the stretched coordinates ($\bxi_1$, $y_2$), the incident wave takes the form
%of $\exp\i (k_1 \bxi_1+ k_2 y_2)  \right)$. Along the edge, the pressure
%variation is therefore T
%For example, the $n$th-order scattered plane
%wave has a form $\exp[\i (k_1 \bxi_1 + (k_2 - 2n\pi)y_2)]$, and therefore the
%pressure variation along the edge of the left teeth as $\bxi_1$ varies from
%$(-1,1)$ is $\exp[\i (k_1 \bxi_1 + (k_2 - 2n\pi)y_2)]$. 

We see from (\ref{equ:FinalFraction}) that the presence of the serration
introduces a modulated streamwise wavenumber of $k_1\pm n\pi/2\bar{h}$ in the
solution. Physically, this can be understood as follows. The presence of the
periodic serration modulates the wavenumber of the incoming plane wave. The
incoming plane wave has a spanwise wavenumber of $k_2$, consequently the $n$th
mode of the scattered pressure has a spanwise wavenumber of $k_2 - 2 n \pi$,
where $n$ is an integer. Because the serration extends in both $y_1$ and $y_2$
directions, the scattered pressure along the edge varies in both $y_1$ and
$y_2$ directions. Therefore, the streamwise wavenumber must be modulated
simultaneously in a similar way as the spanwise wavenumber. Because the half
wavelength (one tooth) and root-to-tip amplitude of the serration are $1/2$ and
$2h$, respectively, the $n$th-order plane wave would have corresponding
modulated streamwise wavenumbers of $k_1+ n\pi/2h$ and $k_1- n\pi/2h$, due to
the presence of the left and right teeth, respectively. Therefore, each
spanwise mode ($n$th for example) corresponds to two equally weighted plane
waves, one with a streamwise wavenumber of $k_1+n\pi/2h$ and the other with
$k_1-n\pi/2h$. We can define two geometrical angles representing the effective
incident angles in the $y_1/\beta-y_3$ plane for the two plane waves, i.e.
\begin{equation}
    \Theta_n^\pm = \arccos \frac{k_1 \pm n \pi/2\bar{h}}{\kappa_n},
    \label{equ:Thetapm}
\end{equation}
where $n$ is an integer. Clearly, for $n=0$ both $\Theta_0^+$ and $\Theta_0^-$
reduce to $\Theta_0 \equiv \arccos k_1/\kappa_0$ representing the incident
angle of $p_{in}$. 

\begin{figure}
    \centering
    \includegraphics[width=0.8\textwidth]{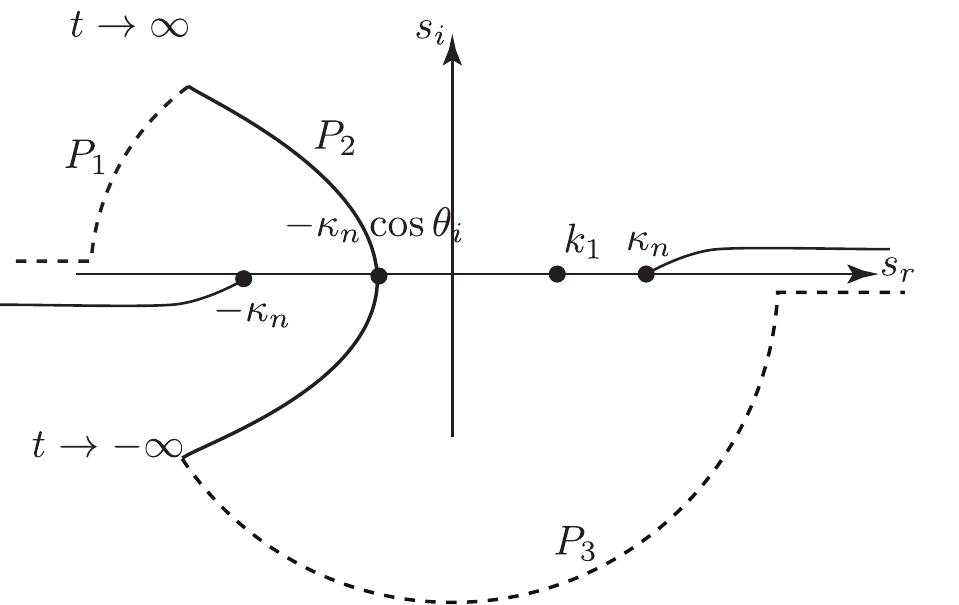}
    \caption{The deformed path $P_1+P_2+P_3$, where $P_2$ is described by the
    $s = -\kappa_n \cos(\theta_i+\i t)$ as $t$ varies from $+\infty$ to
    $-\infty$. When $t=0$ the path $P_2$ intersects with the real axis at
    $-\kappa_n\cos\theta_i$. When $-\kappa_n\cos\theta_i > k_1$ as shown above,
    the simple pole is crossed, and a residue contribution must be included.
    The case when $\kappa_n$ is imaginary is similar.}
    \label{fig:deformedPath}
\end{figure}
To evaluate (\ref{equ:FinalFraction}), we deformed the integration path
$P_0$ shown in figure~\ref{fig:integralPath} to the curve $P_1+P_2+P_3$ shown
in figure~\ref{fig:deformedPath}. The path $P_2$ is described by $s= -\kappa_n
\cos(\theta_i + \i t)$, where the real number $t$ varies from $+\infty$ to
$-\infty$. Figure~\ref{fig:deformedPath} shows that $s\to \infty$ in the second
quadrant as $t \to +\infty$, whereas as  $s\to \infty$ in the third quadrant $t
\to -\infty$. It can be shown that integration along $P_1$ and $P_3$ approaches
$0$ as $|s|\to\infty$. Therefore the integral in
(\ref{equ:FinalFraction}) can be evaluated along $P_2$ instead provided
$P_2$ pass the simple pole from below. Because the deformed path $P_2$
intersects with the real axis at $-\kappa_n \cos\theta_i$, such condition is
met when $-\kappa_n \cos\theta_i < k_1$, i.e. when $\theta_i >
\pi-\arccos(k_1/\kappa_n)$. When $0<\theta_i < \pi - \arccos(k_1/\kappa_n)$, we
can show that a residue contribution must be added. However, this pole
contribution is exactly cancelled by the jump in the resulting integral and the
final solution takes the same form as that for $\theta_i > \pi - \arccos(k_1
/\kappa_n)$ (see Chapter 2 of \cite{Noble1958} for detail). Therefore, in the
rest of the paper we choose not to distinguish the two cases. By deforming the
integral along $P_2$ and making use of the definition of
(\ref{equ:Thetapm}), (\ref{equ:FinalFraction}) reduces to
\begin{equation}
    D_n^\pm(r_i, \theta_i) 
    = -\sqrt{\frac{2}{\kappa_n}}
    \int_{-\infty}^{\infty}
    \frac{\sin\frac{1}{2}(\theta_i + \i t)}
    {\cos(\theta_i+\i t) + \cos\Theta_n^{\pm}}
    \e^{\i \kappa_n r_i \cosh t} \ud t.
	\label{equ:ThetaIntegration}
\end{equation}
Equation~(\ref{equ:ThetaIntegration}) can be integrated analytically to yield
(see Appendix~\ref{app:HnEvaluation} for more details) 
\begin{equation}
    D_n^\pm(r_i, \theta_i) =
    -\pi\sqrt{\frac{2}{\kappa_n}}
     \frac{I(\kappa_n r_i, \theta_i; \Theta_n^\pm)}{\sin
     \frac{1}{2}\Theta_n^\pm},
     \label{equ:Hpm}
\end{equation}
where function $I(kr, \theta; \Theta)$ is the classical
Fresnel solution denoting the pressure field scattered by a straight trailing
edge~\citep{Noble1958}, i.e.
\begin{equation}
    I(kr, \theta; \Theta) = \frac{\e^{-\i \frac{\pi}{4}}}{\sqrt{\pi}}  
    \left[
	\e^{-\i kr \cos(\Theta+\theta)} F(\sqrt{2kr} \cos
	\frac{\Theta+\theta}{2})
	-
	\e^{-\i kr \cos(\Theta-\theta)} F(\sqrt{2kr} \cos
	\frac{\Theta-\theta}{2})
    \right].
    \label{equ:standardIn}
\end{equation}
The Fresnel integral $F(x)$ in (\ref{equ:standardIn}) is defined as
\begin{equation}
    F(x) = \int_x^\infty \e^{\i u^2} \ud u
    \label{equ:FresnelIntegral}
\end{equation}
and can be conveniently computed using the standard error function. 

Having obtained the analytical result of $D_n^\pm(r_i, \theta_i)$, it follows
that
\begin{equation}
    H_n(r_i, \theta_i) = \frac{\i}{\sqrt{2\kappa_n}n}
    \left( 
	\frac{I(\kappa_n r_i, \theta_i; \Theta_n^+)}{\sin\frac{1}{2}\Theta_n^+}
	-\frac{I(\kappa_n r_i, \theta_i; \Theta_n^-)}{\sin\frac{1}{2}\Theta_n^-}
    \right) 
    \label{equ:Hfun}
\end{equation}
where as mentioned above the subscript $i$ take the value of either $t$ or $r$.
Note that in (\ref{equ:Hfun}), $H_n(r_i, \theta_i)$ decays at least as fast as
$n^{-3 / 2}$ as $n\to \infty$, and because $n$ appears in the denominator,
(\ref{equ:Hfun}) works only for $n\ne 0$. However, if treating $n$ as real
variable, we may obtain the result for $n=0$ by taking the limit as $n\to 0$.
To facilitate practical computations, we also derive an explicit formula for
$H_0(r_i, \theta_i)$ from (\ref{equ:Hn}). This can be found in
Appendix~\ref{app:H0Integration}.

Substituting (\ref{equ:Hfun}) into (\ref{equ:Gs}), the total scattered pressure
$G_s$ can be readily evaluated. The important fact is that (\ref{equ:Gs}) is an
exact evaluation of (\ref{equ:GsIntegral}), and therefore is not only valid in
the far field, but also in the near field. When $r\to \infty$, (\ref{equ:Gs})
would recover the far-field approximation obtained using the steepest descent
method by \citet{Ayton2018d}. Note again that (\ref{equ:Hfun}) consists of the
standard Fresnel solution describing the scattered field by a straight edge.
This suggests that the pressure field scattered by the sawtooth edge is
equivalent to the sum of the Floquet modes scattered by two imagined
semi-infinite flat plates with their straight trailing edges located at the tip
and root of the serration, respectively. This appears to be somewhat consistent
with a number of previous findings showing that noise generation by serrated
edges is dominated by the root or tip
regions~\citep{Kim2016,Turner2017,Avallone2018}. This
view, however, results from the use of the $E_n(s)$ assumption and is therefore
not exact. Because (\ref{equ:EnSawtooth}) is used in the derivation,
(\ref{equ:Gs}) is therefore only valid for sawtooth serrations; however, as
mentioned earlier we can easily obtain analytical Green's functions for any
arbitrary piecewise linear serration profiles. Appendix~\ref{app:OtherGs}
contains the Green's functions for other serration profiles, such as the square
shapes.

When a point source is located at $\boldsymbol{x}$, i.e. $(x_1,
x_2, x_3)$, the incident plane wave near the serration has an amplitude of 
\begin{equation} 
    A(\boldsymbol{x}) = -\frac{1}{\beta} \frac{1}{4 \pi R} \e^{\i k R / \beta} 
    \e^{-\i\frac{kM}{\beta^2}x_1},
\end{equation}
where $R = \sqrt{(x_1/\beta)^2 + x_2^2 + x_3^2}$. Furthermore, the value of
$k_1$ and $k_2$ in the definition of $p_{in}$ can be found to be 
\begin{equation}
    k_1 = \frac{k}{\beta} \frac{x_1 /\beta}{R}, \quad k_2 = \frac{k}{\beta}
    \frac{x_2}{R}.
\end{equation}
By linearity, the Green's function can be readily obtained as
\begin{equation}
    G(\boldsymbol{x}; \boldsymbol{y},\omega) 
    = G^a(\boldsymbol{y}; \boldsymbol{x}, \omega) 
    = A(\boldsymbol{x}) \left(p_{in} + G_s\right), 
\label{equ:analyticalGreensFunction}
\end{equation}
where $p_{in}$ is shown in (\ref{equ:incidentWave}) and $G_s$ is given
by (\ref{equ:Gs}).
Equation~(\ref{equ:analyticalGreensFunction}) is the fundamental equation of this
paper. It can be seen that the Green's function consists of two terms; the
first term $A(\boldsymbol{x}) p_{in}$ represents the sound propagating directly
from the source $\boldsymbol{y}$ to the observer $\boldsymbol{x}$, while the
second term $A(\boldsymbol{x})G_s$ represents the scattered pressure off the
serrated plate then propagating to the observer $\boldsymbol{x}$. The direct
propagating sound is trivial and most importantly does not depend on the
serration profiles, therefore it is the scattered part that we are interested
in.

\section{Validation}
\label{sec:Validation}
We see from section~\ref{sec:GreensFunction} that in order to obtain the
analytical Green's function, considerable algebra is involved. Therefore, it is
necessary to validate the result before the Green's function is used to study
the scattering characteristics. In this section, we choose to validate the
Green's function using two approaches. The first is to numerically integrate
(\ref{equ:GsIntegral}) so as to ensure that the complex analytical
evaluation of the integral is correct. The second approach is to make use of
the FEM technique to compute the scattered pressure under the incident wave
shown in (\ref{equ:incidentWave}) using COMSOL so as to examine to which
extent the assumption regarding $E_n(s)$ serves as a good approximation. 
\begin{figure}
    \centering
    \includegraphics[width=0.9\textwidth]{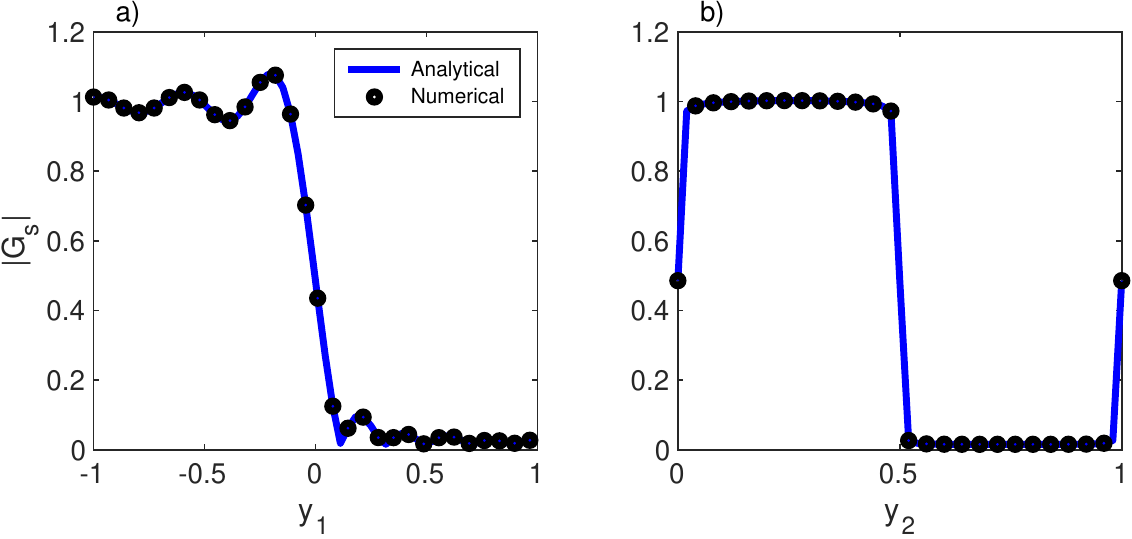}
    \caption{Comparison of the scattered pressure field $|G_s|$ on (a) $y_2=0$,
    $y_3=0$ and (b) $y_1=0$ and $y_3 = 0$. The serration amplitude is $h=10$,
    the wavenumber is $k=1$, the Mach number $M=0$ and the observer angle
    $\Theta_0=\frac{3}{4}\pi$.}
    \label{fig:numericalIntegration}
\end{figure}

Figure~\ref{fig:numericalIntegration} shows a comparison between the scattered
pressure $|G_s|$ obtained by the numerical integration and from
(\ref{equ:Gs}). As can be seen from
figure~\ref{fig:numericalIntegration}(a), the scattered pressures obtained
using the two approaches completely collapse along the line of $y_2 = 0 $ in
the plane of the flat plate. Similarly, the scattered pressure obtained by
numerical integration along the line of $y_1=0$ in the flat plate plane is
identical to that from (\ref{equ:Gs}). Pressure values at other
locations show exactly the same agreement. This excellent agreement shows that
the analytical evaluation of the contour integral in the complex $s$ domain is
indeed correct and exact. 

%One important fact about
%figure~\ref{fig:numericalIntegration}(a) is that the scattered pressure
%$G_s(s)$ is not strictly equal to $0$ for for $y>0$, although the boundary
%conditions shows that it should be. Indeed, this is due to the outcome of
%assuming $E_n(s)$ is a factor of $R_n^-(s)$ and $R_n^{+\prime}(s)$. However,
%from figure~\ref{fig:numericalIntegration}(a-b) we see that the boundary
%condition is approximated reasonably well.

\begin{figure}
    \centering
    \includegraphics[width=0.8\textwidth]{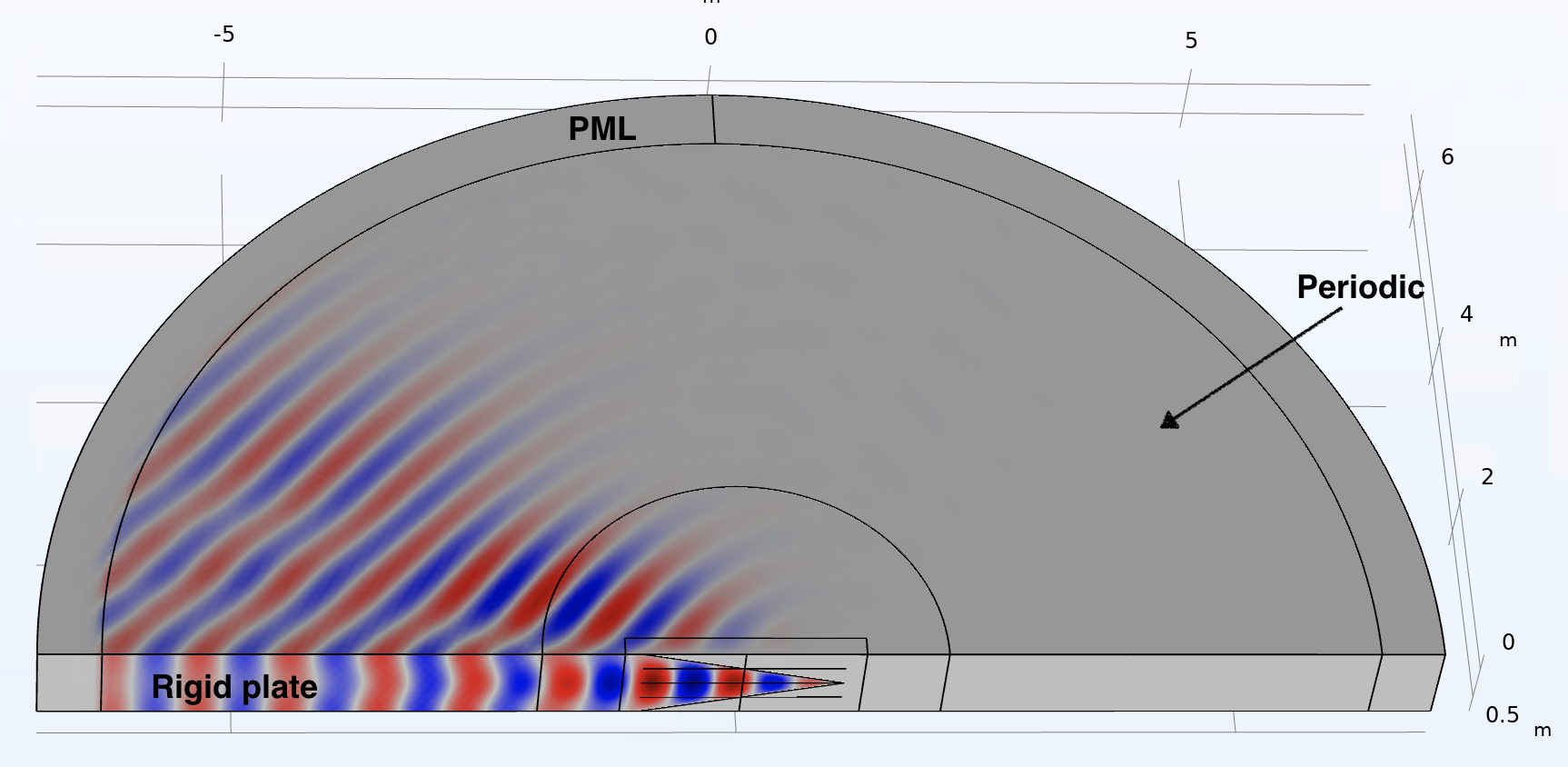}
    \caption{Scattered pressure field $G_s$ (only real part is shown) from FEM
    simulations, where $M=0$, $h=2$, $k=10$ and $\Theta=\pi/4$. A small
    imaginary part of $k$ is used to improve the PML accuracy.}
    \label{fig:FEM-Setup}
\end{figure}
Although figure~\ref{fig:numericalIntegration} shows that the analytical
derivation from (\ref{equ:GsIntegral}) to (\ref{equ:Hfun}) is correct, it
cannot show  to what extent (\ref{equ:analyticalGreensFunction}) approximates
the exact solution to (\ref{equ:ConvectiveWaveEquation}). This is because
(\ref{equ:GsIntegral}) is based on the assumption of $E_n(s)$, and to examine
its validity we need to use FEM to numerically calculate the scattered pressure
so that a direct comparison between the numerical and analytical Green's
functions can be made. We again choose to compare the near-field $|G_s|$ under
the incident wave shown in (\ref{equ:incidentWave}). 

The commercial software COMSOL is used to conduct the numerical simulation, the
computational domain of which is shown in figure~\ref{fig:FEM-Setup}. We can
see that a half-cylindrical domain consisting of one serration wavelength is
used. The semi-infinite plate is placed on the left-hand side of the bottom
surface, as shown in figure~\ref{fig:FEM-Setup}. Periodic boundary conditions
are used between the front and back surfaces. Perfectly Matched Layers (PML)
are attached to the outer side of the domain to absorb the scattered pressure
due to a plane wave incidence prescribed by (\ref{equ:incidentWave}). The PML
works well for absorbing sound scattered off a finite object, but starts to
become less accurate to simulate a flat plate that is semi-infinitely long. To
improve the accuracy of the PML, a small imaginary part of $k$ (in this paper
$\arg{k}\approx -0.02$) is used so that the scattered pressure decays gradually
as it propagates. When compared against analytical results, the same $k$ is
used in (\ref{equ:Gs}). This is permissible and \add{can be shown conveniently
by analytical continuation.} \add{Free tetrahedral mesh is used} and the
resulting case has up to 4 millions degrees of freedom at the highest
dimensionless frequency $k$. Grid independence is examined by using
increasingly fine meshes that result in little change in the calculated
pressure field.

\begin{figure}
    \centering
    \includegraphics[width=0.8\textwidth]{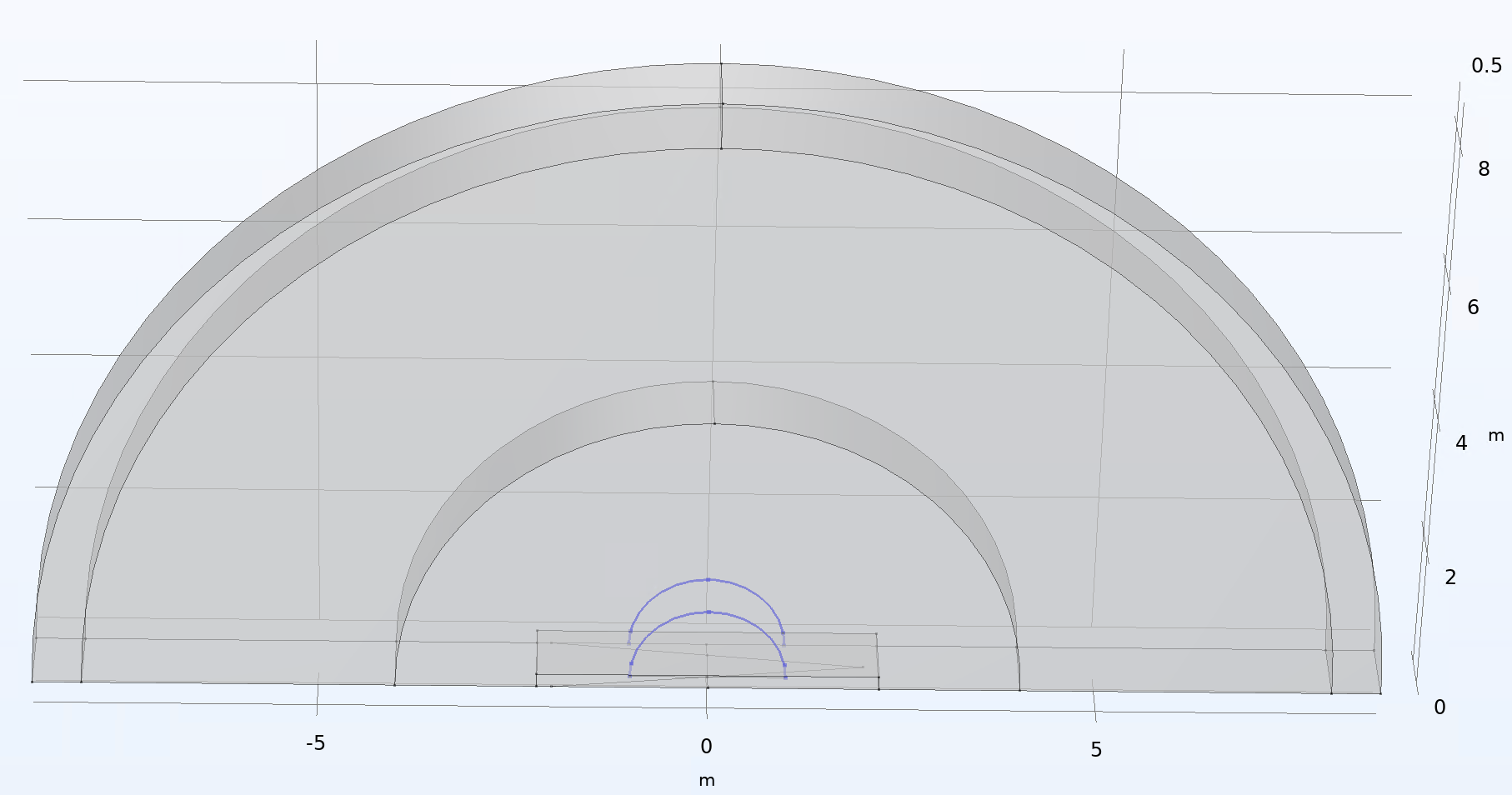}
    \caption{Two semicircles of radius $1$ denote the probe locations on which
    two scattered pressure are compared between the FEM and the analytical
    formula. The front semicircle is at $y_2=0$ while the back one is at $y_2 =
    0.75$.} 
    \label{fig:ProbeLocation}
\end{figure}

The scattered near-field pressure is evaluated along two semicircles shown in
blue in figure~\ref{fig:ProbeLocation}. The two semicircles have a
dimensionless radius of $1$ and are located in the $y_2=0$ and $y_2 =0.75$
planes, respectively. In the rest of this paper, they are referred to as the SC
1 ($y_2=0$) and SC 2 ($y_2 = 0.75$), respectively. In the FEM computation, the
serration amplitude $h$, the frequency $k$ and the incident angle $\Theta$ can
all be varied. To facilitate comparison, the scattered pressure by a straight
trailing edge is also computed and evaluated on the same semicircles. 

\begin{figure}
    \centering
    \begin{subfigure}{0.47\textwidth}
	\centering
	\includegraphics[width=\textwidth]{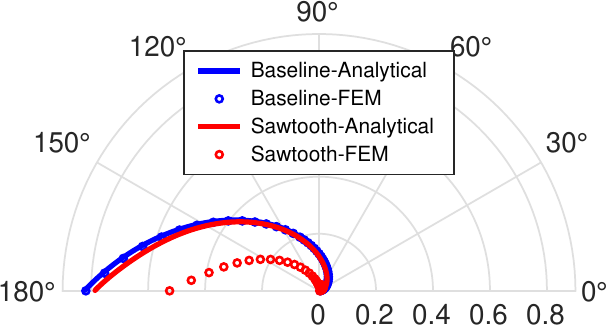}
	\caption{$k=2$, SC 1}
    \end{subfigure}
    \begin{subfigure}{0.47\textwidth}
	\centering
	\includegraphics[width=\textwidth]{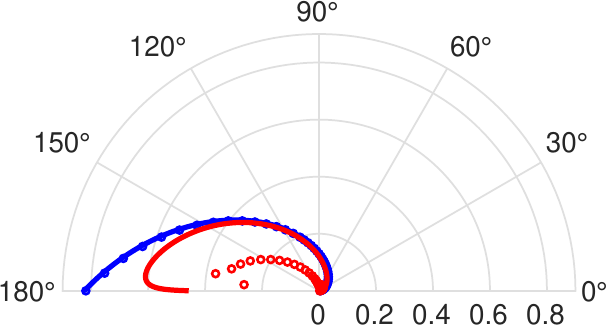}
	\caption{$k=2$, SC 2}
    \end{subfigure}
    \begin{subfigure}{0.47\textwidth}
	\centering
	\includegraphics[width=\textwidth]{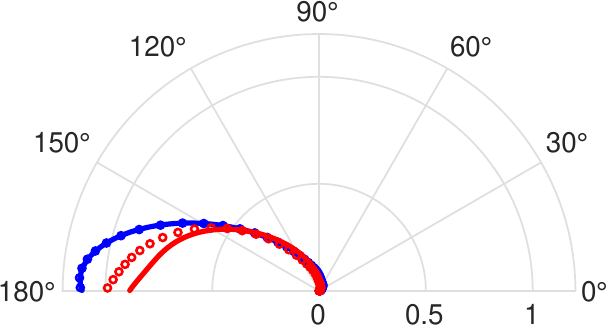}
	\caption{$k=10$, SC 1}
    \end{subfigure}
    \begin{subfigure}{0.47\textwidth}
	\centering
	\includegraphics[width=\textwidth]{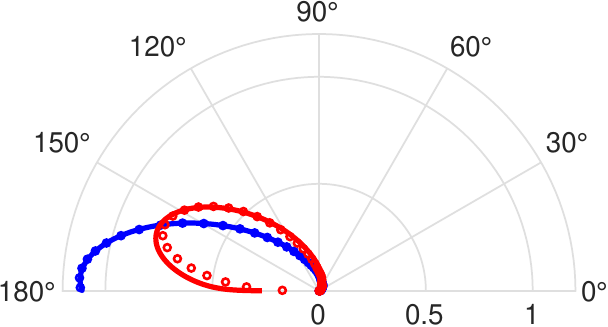}
	\caption{$k=10$, SC 2}
    \end{subfigure}
    \begin{subfigure}{0.47\textwidth}
	\centering
	\includegraphics[width=\textwidth]{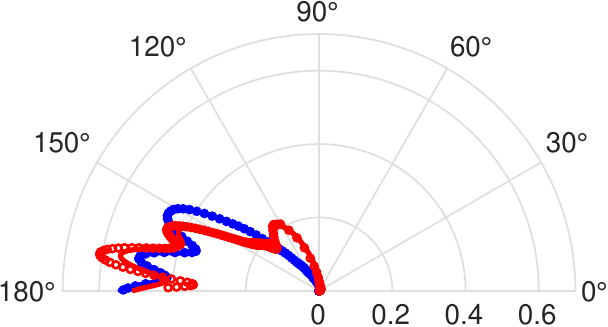}
	\caption{$k=50$, SC 1}
    \end{subfigure}
    \begin{subfigure}{0.47\textwidth}
	\centering
	\includegraphics[width=\textwidth]{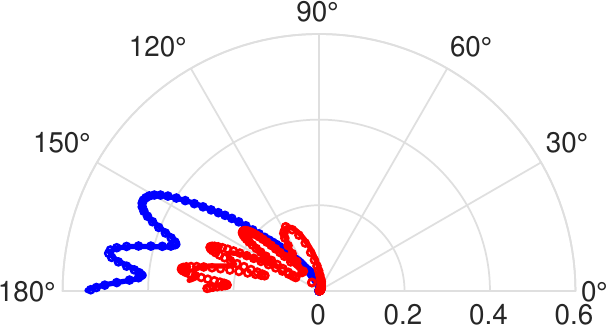}
	\caption{$k=50$, SC 2}
    \end{subfigure}
    \caption{Comparison of the analytical and FEM-calculated Green's function
    at $M = 0$ and $\Theta=\pi / 4$ for both baseline and serrated ($h=1$)
    trailing edges on the first (SC 1) and second (SC 2) semicircles at various
    frequencies.}
    \label{fig:FEMTheta1/4Pi}
\end{figure}

Figure~\ref{fig:FEMTheta1/4Pi} shows the comparison of the scattered pressure
calculated by the FEM technique and the analytical Green's function at $M=0$
and $\Theta=\pi/4$. The scattered pressure values from both the straight (blue)
and serrated (red) edges at various non-dimensional frequencies are shown.
Figure~\ref{fig:FEMTheta1/4Pi}(a-b) shows the results when $k=2$, from which we
can see that the computed pressure distribution for straight edges agrees
excellently with the analytical prediction. This ensures that the PML works
satisfactorily and the grid is sufficiently fine to resolve the pressure field.
On the other hand, the computed scattered pressure for serrated edges is
significantly smaller than the analytical prediction. This is expected, as the
assumption about $E_n(s)$ dependence is not expected to be valid at this
frequency. As mentioned in section~\ref{sec:GreensFunction}, as the frequency
increases the scattering becomes increasingly localized and the assumption is
more likely to be valid. This is indeed the case, as shown in
figure~\ref{fig:FEMTheta1/4Pi}(c-d). We can see that the baseline results
continue to agree excellently with analytical predictions, but the serrated
results are now in better agreement with the analytical prediction with slight
deviation at large angles. In particular, figure~\ref{fig:FEMTheta1/4Pi}(d)
shows a significant change in the near-field directivity shape on SC2 due to
the use of serrations, and the Green's function can capture this change well
apart from the small magnitude deviation at large observer angles. As the
frequency increases to $k=50$, figure~\ref{fig:FEMTheta1/4Pi}(e-f) shows that
the scattered pressure obtained using the two methods agree well with each other
both in terms of the shape and amplitude of the directivity patterns. 

\begin{figure}
    \centering
    \begin{subfigure}{0.47\textwidth}
	\centering
	\includegraphics[width=\textwidth]{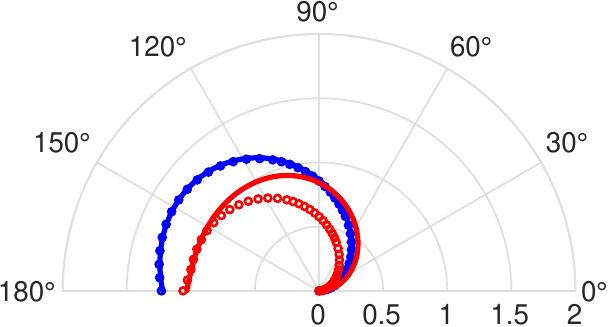}
	\caption{$k=2$, SC 1}
    \end{subfigure}
    \hfill
    \begin{subfigure}{0.47\textwidth}
	\centering
	\includegraphics[width=\textwidth]{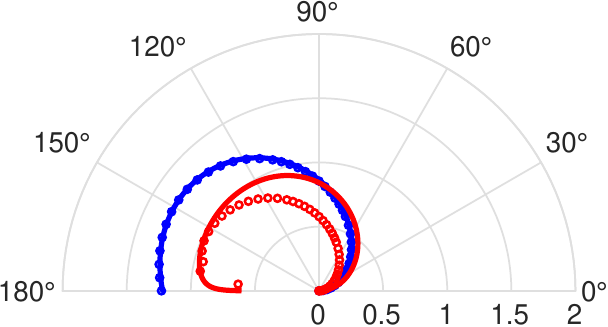}
	\caption{$k=2$, SC 2}
    \end{subfigure}
    \begin{subfigure}{0.47\textwidth}
	\centering
	\includegraphics[width=\textwidth]{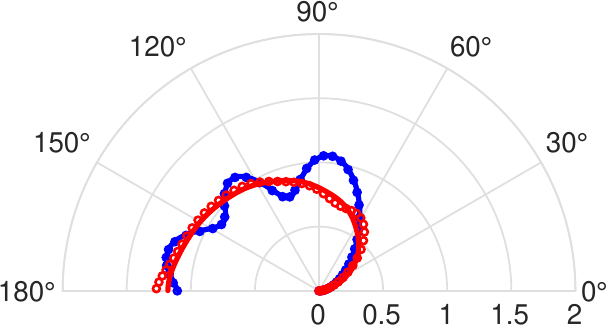}
	\caption{$k=10$, SC 1}
    \end{subfigure}
    \hfill
    \begin{subfigure}{0.47\textwidth}
	\centering
	\includegraphics[width=\textwidth]{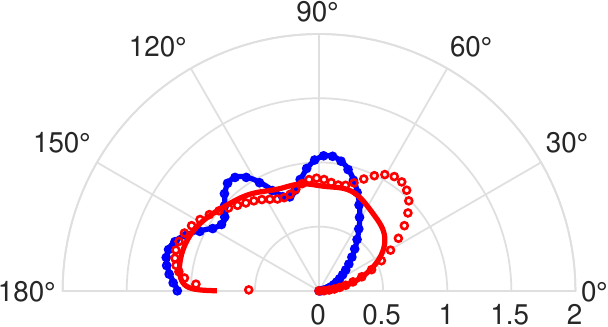}
	\caption{$k=10$, SC2}
    \end{subfigure}
    \begin{subfigure}{0.47\textwidth}
	\centering
	\includegraphics[width=\textwidth]{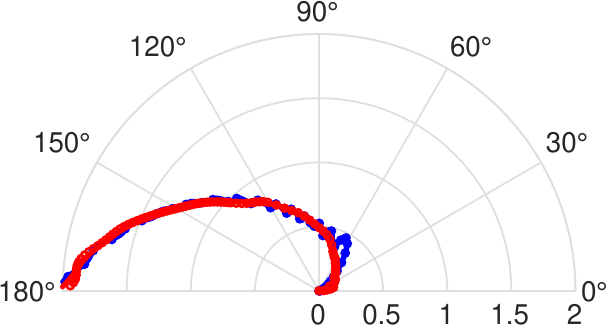}
	\caption{$k=50$, SC 1}
    \end{subfigure}
    \hfill
    \begin{subfigure}{0.47\textwidth}
	\centering
	\includegraphics[width=\textwidth]{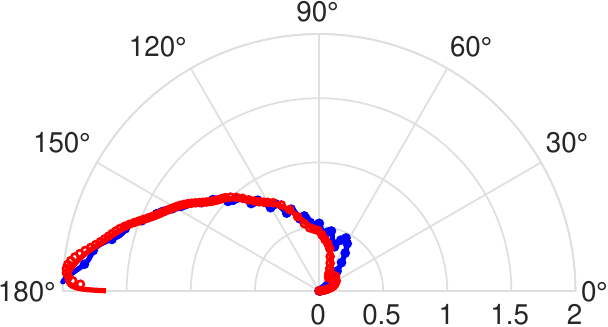}
	\caption{$k=50$, SC2}
    \end{subfigure}
    \caption{Comparison of the analytical and FEM-calculated Green's function
    at $M = 0$ and $\Theta=3\pi/4$ for both baseline and serrated ($h=1$)
    trailing edges on the first (SC 1) and second (SC 2) semicircles at various
    frequencies. \add{Legends are the same as those shown in
figure~\ref{fig:FEMTheta1/4Pi}.}}
    \label{fig:FEMTheta3/4Pi}
\end{figure}

Figure~\ref{fig:FEMTheta3/4Pi} shows the comparison between the scattered
pressure when the incident angle $\Theta=3/4\pi$ for both the baseline and
serrated trailing edges. When the incident angle $\Theta=3/4\pi$, the
directivity patterns of the scattered pressure are significantly different from
those at $\Theta=\pi/ 4$. Nevertheless, figure~\ref{fig:FEMTheta3/4Pi}(a-b)
still clearly shows the discrepancy between the FEM and analytical results for
serrated edges, while figure~\ref{fig:FEMTheta3/4Pi}(c-d) shows that the
agreement is fairly good. Again at the highest frequency $k=50$, the two lines
virtually collapse, as shown in figure~\ref{fig:FEMTheta3/4Pi}(e-f),
indicating good agreement between the analytical and calculated Green's
functions.

In summary, we see that the analytical Green's function approximates the exact
Green's function reasonably well at high frequencies. A rule of thumb for the
valid regime may be taken as $kh > 10$. It is, however, worth noting that the
incident plane wave is by no means limited to propagative waves. Because of
analytical continuation, the Green's function must also work for evanescent
waves, such as the plane-wave gusts used in TE noise modelling using Amiet's
approach. In such cases, the convective Mach number of the gust is typically
low and therefore it is the hydrodynamic wavenumber $k_1h$ that determines how
localized the scattering is. Considering that this hydrodynamic wavenumber is
often much larger than the acoustic wavenumber, especially for low Mach number
applications, the $E_n(s)$ assumption would be more likely to hold so that
(\ref{equ:Gs}) may be used to develop a three-dimensional TE noise
prediction model. Note that although (\ref{equ:Gs}) appears complex, it
can be simplified considerably when used to model TE noise because both
$\Theta$ and $\theta_i$ are equal to $\pi / 2$ in the scattered surface
pressure calculation. Therefore, it can be expected the resulting model can be
cast into a relatively compact form that facilitates efficient evaluation.

\section{Results and Discussions}
\label{sec:Results}
Having validated the analytical Green's function, we are now in a position to
examine its far-field radiation characteristics and the effects of varying the
frequency, serration amplitude, sound source position and Mach number
respectively using (\ref{equ:analyticalGreensFunction}). As shown in
section~\ref{sec:GreensFunction}, (\ref{equ:analyticalGreensFunction})
consists of a directly propagating incident part and a scattered part. In the
rest of this section, we only examine the scattered part of Green's function,
i.e. $A(\boldsymbol{x})G_s$ shown in (\ref{equ:Gs}), because it is the
scattered part $A(\boldsymbol{x})G_s$ that is related to the serration
geometry, whereas the incident part remains unchanged no matter how the
serration changes. To study the effects of the frequency, the
non-dimensionalized wavenumbers $k=10$ and $k=50$ are chosen because
section~\ref{sec:Validation} shows that the Green's function serves as a
reasonably good approximation to the exact solution at these frequencies. In
the rest of this section, the far-field directivity patterns at $k=10$ and
$k=50$ are shown simultaneously when the serration amplitude, source position
and Mach number vary. To account for the pressure decay due to sound
propagation, the scattered pressure in
(\ref{equ:analyticalGreensFunction}) is scaled by $4\pi|\boldsymbol{x}|$
so that
\begin{equation}
    G_{scaled} = -\frac{1}{\beta} \frac{|\boldsymbol{x}|}{R} \e^{\i k R / \beta} 
    \e^{-\i\frac{kM}{\beta^2}x_1}G_s
    \label{equ:Gscaled}
\end{equation}
is used to plot the directivity patterns. In the following directivity plots,
the observer position is chosen to be $|\boldsymbol{x}|=100$ away from the
coordinate origin in the $x_2= 0$ plane. From (\ref{equ:Gscaled}), we see that
$|G_{scaled}| = |G_s||\boldsymbol{x}|/\beta R$, therefore choosing values other
than $100$ does not change the directivity shape and magnitude in the plane
$x_2=0$.

\subsection{Effects of the serration amplitude} 
\begin{figure}
    \centering
    \begin{subfigure}{0.47\textwidth}
    \centering
    \includegraphics[width=\linewidth]{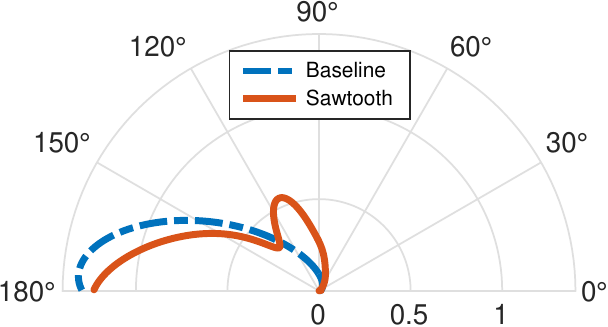}
    \caption{$h=1, k=10$}
    \end{subfigure}
    \hfill
    \begin{subfigure}{0.47\textwidth}
    \centering
    \includegraphics[width=\linewidth]{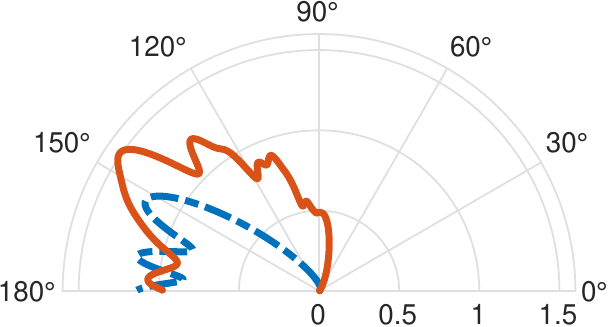}
    \caption{$h=1, k=50$}
    \end{subfigure}
    \begin{subfigure}{0.47\textwidth}
    \centering
    \includegraphics[width=\linewidth]{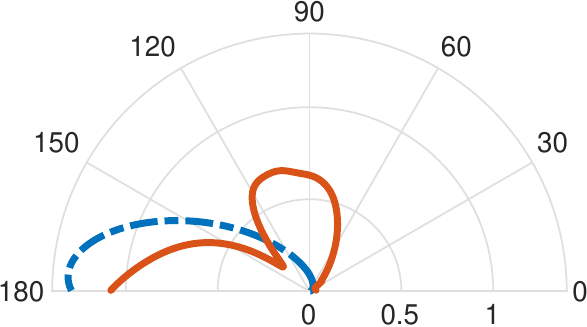}
    \caption{$h=2, k=10$}
    \end{subfigure}
    \hfill
    \begin{subfigure}{0.47\textwidth}
    \centering
    \includegraphics[width=\linewidth]{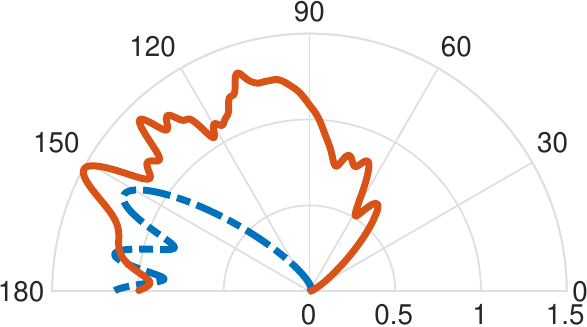}
    \caption{$h=2, k=50$}
    \end{subfigure}
    \begin{subfigure}{0.47\textwidth}
    \centering
    \includegraphics[width=\linewidth]{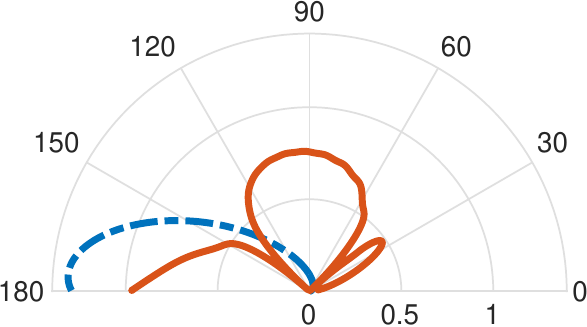}
    \caption{$h=5, k = 10$}
    \end{subfigure}
    \hfill
    \begin{subfigure}{0.47\textwidth}
    \centering
    \includegraphics[width=\linewidth]{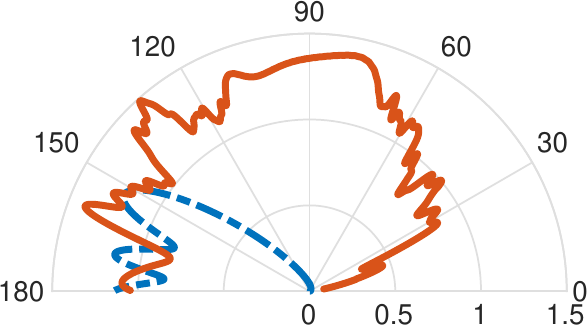}
    \caption{$h=5, k = 50$}
    \end{subfigure}
    \caption{Far-field directivity patterns of the scattered pressure due to
    the baseline and serrated edges of different sizes. The source is located
    at $r=1$ and $\theta = \pi/4$ in the $y_2=0.25$ plane (passing the
    serration tip).}
    \label{fig:AmplitudeEffects}
\end{figure}

We first study the effects of varying the serration amplitude on the scattering
characteristics. As a starting point we let $M=0$ so that $\beta=1$, i.e. the
stretched coordinates are just the physical coordinates in the definition of
$G_s$. We choose three serration lengths, i.e. $h=1$, $h=2$ and $h=5$, and have
their scattered far-field directivity $|G_{scaled}|$ plotted in
figure~\ref{fig:AmplitudeEffects}. The directivity plots for their
corresponding straight-edge scattering are also included for reference. In all
these plots we fix the source position at $(\cos\pi/4, 0.25, \sin\pi/4)$. From
figure~\ref{fig:GeoAngleDef} it can be seen that $y_2=0.25$ is the plane that
passes the serration tip. From the rest of the paper, we refer to it as the tip
plane, whereas $y_2=0.75$ similarly referred to as the root plane. 
Figure~\ref{fig:AmplitudeEffects}(a-b) shows the directivity patterns for a
serration amplitude of $h = 1$. \add{This represents a relatively wide
serration. Note we do not consider serrations wider than this,} i.e. serrations
with very small $h$ values, because that would result in very small $kh$ values
that invalidate the $E_n(s)$ assumption. This however does not pose much
restriction on its applications, because it is widely known that serrations
with very short amplitude have little effect on reducing TE noise. We see from
figure~\ref{fig:AmplitudeEffects}(a) that compared to the baseline results, the
scattered pressure is slightly weaker at large observer angles, i.e. $\Theta >
135^\circ$, but slightly stronger at others. As the frequency increases to
$k=50$, we see a more pronounced radiation enhancement when $\Theta <
135^\circ$ and an increasingly less obvious noise suppression at large observer
angles. Besides, both the baseline and serrated directivity patterns exhibit
lobes resulting from interference between the geometrically-reflected and
scattered pressure fields.

\begin{figure}
\centering
\begin{subfigure}{0.47\textwidth}
\centering
\includegraphics[width=\linewidth]{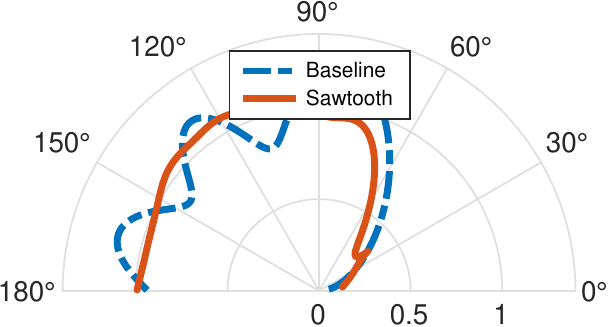}
\caption{$h=1$, $k=10$}
\end{subfigure}
\hfill
\begin{subfigure}{0.47\textwidth}
\centering
\includegraphics[width=\linewidth]{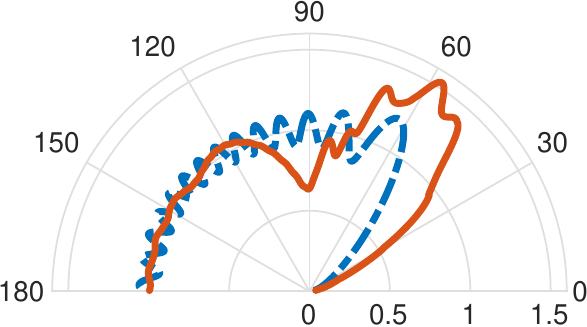}
\caption{$h=1$, $k=50$}
\end{subfigure}
\begin{subfigure}{0.47\textwidth}
\centering
\includegraphics[width=\linewidth]{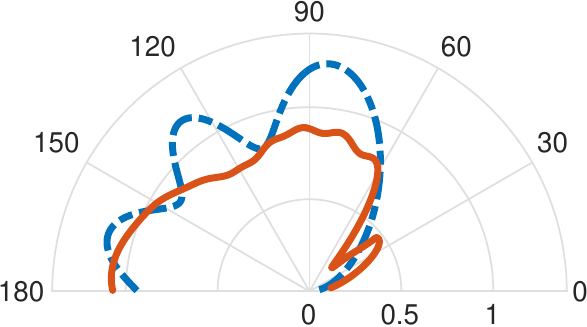}
\caption{$h=2$, $k=10$}
\end{subfigure}
\hfill
\begin{subfigure}{0.47\textwidth}
\centering
\includegraphics[width=\linewidth]{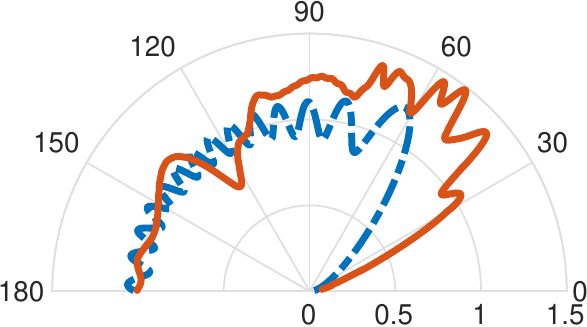}
\caption{$h=2$, $k=50$}
\end{subfigure}
\begin{subfigure}{0.47\textwidth}
\centering
\includegraphics[width=\linewidth]{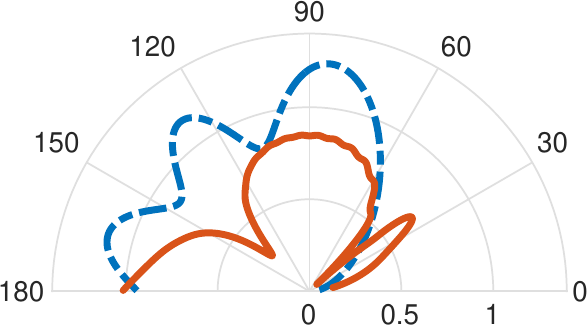}
\caption{$h=5$, $k = 10$}
\end{subfigure}
\hfill
\begin{subfigure}{0.47\textwidth}
\centering
\includegraphics[width=\linewidth]{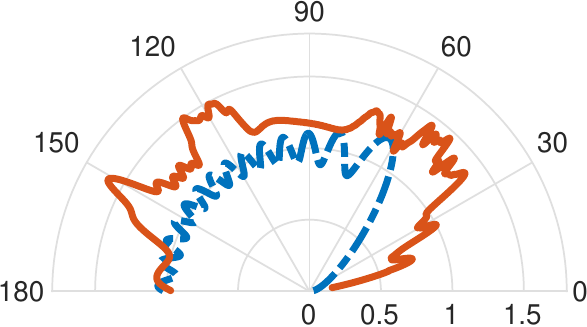}
\caption{$h=5$, $k = 50$}
\end{subfigure}
\caption{Far-field directivity patterns of the scattered pressure due to the
baseline and serrated edges of different sizes. The source is located at $r=1$
and $\theta = 3\pi/4$ in the $y_2=0.25$ plane (the serration tip plane).}
\label{fig:AmplitudeEffects2}
\end{figure}

As the amplitude of the serration increases, this tendency becomes increasingly
evident, as shown in figure~\ref{fig:AmplitudeEffects}(c-d), where directivity
patterns for serrations of $h=2$ are shown.
Figure~\ref{fig:AmplitudeEffects}(e-f) shows the directivity patterns when $h=
5$. This represents a rather long serration, and we see that the general
behaviour of the scattered pressure remains similar to serrations of $h=2$. The
difference is that the low-angle enhancement is more evident, in particular
when $\Theta<135^\circ$. For the long serration shown in
figure~\ref{fig:AmplitudeEffects}(e-f),  we see that the directivity shapes are
significantly different from that of straight edges. This may be understood as
follows. When the amplitude of the serration increases, the additional area of
the serration extended downstream can act as an effective reflection surface
for the nearby source when the observer is at low observer angles. From the
directivity plots it manifests itself as large noise radiation or even an
additional lobe, as shown in figure~\ref{fig:AmplitudeEffects}(a-f). For
example, because the sound source is at $\theta=\pi/4$ in the tip plane in
figure~\ref{fig:AmplitudeEffects}, the geometrically reflected acoustic wave
would only exist in the range of $135^\circ < \Theta \le  180$ for the baseline
flat plate. The scattered pressure gradually decreases to $0$ as $\Theta$
reduces to $0$. However, when a sufficiently large serration exists, the
extended surface would provide additional reflection for a nearby source and
stronger noise radiation would occur at low observer angles (e.g. $\Theta <
135^\circ$) due to the additional reflection. This implies that turbulence
eddies directly above the surface in the tip plane are more efficient in
radiating noise to low observer angles and therefore are of more relevance for
noise suppression. 

The effects of serration amplitude can also be studied when the source is
located at $\theta=3\pi/4$ in the tip plane. The results are shown in
figure~\ref{fig:AmplitudeEffects2}. Figure~\ref{fig:AmplitudeEffects2}(a-b)
shows the directivity patterns for wide serrations of $h=1$. Because the source
is located at $\theta=3\pi/4$, the scattered pressure by the baseline trailing
edge has a large amplitude when $\Theta > 45^\circ$ because of surface
reflection. When the serrated TE is used, similar reflection exists and the
resulting directivity is therefore similar to the baseline results. As the
frequency increases to $k=50$ the behaviour remains largely similar apart from
multiple lobes resulting from the interference. However below $\Theta=60^\circ$
we also see a noticeable noise increase, this again may be explained by the
extended surface downstream of the source. As the serration amplitude further
increases, we expect a more pronounced noise increase. This in fact can be seen
in figure~\ref{fig:AmplitudeEffects2}(c), where a small radiation lobe starts
to appear. At $k=50$ the noise enhancement at $\Theta<60^\circ$ is more
evident. From figure~\ref{fig:AmplitudeEffects2}(e), we see a roughly similar
behaviour for the longest serration of $h=5$. However, we also note a slightly
weaker radiation for $\Theta>60^\circ$. This may be due to the fact that part
of the flat plate upstream of the source is removed, therefore weakening a
perfect reflection by a straight TE. Such weakening, however, can be expected
to diminish as the frequency increases leading to increasingly localized
scattering, which is indeed the case, as shown in
figure~\ref{fig:AmplitudeEffects2}(f). When the serration amplitude is large
and the source is located in the close vicinity of the origin, it can be
expected the scattered directivity would remain roughly similar between a
source at $\theta=\pi/4$ and the other at $\theta=3\pi/4$. By comparing
figures~\ref{fig:AmplitudeEffects}(f) and \ref{fig:AmplitudeEffects2}(f) we see
this is indeed the case.
\begin{figure}
    \centering
    \begin{subfigure}{0.47\textwidth}
    \centering
    \includegraphics[width=\linewidth]{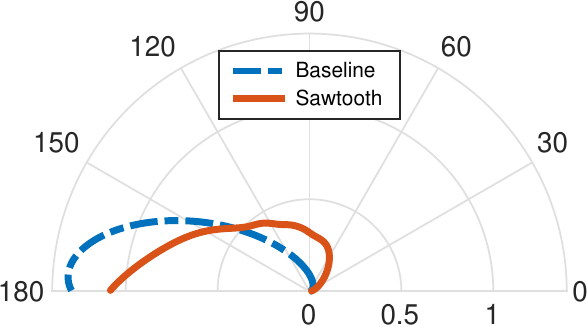}
    \caption{$y_2 = 0$, $k=10$}
    \end{subfigure}
    \hfill
    \begin{subfigure}{0.47\textwidth}
    \centering
    \includegraphics[width=\linewidth]{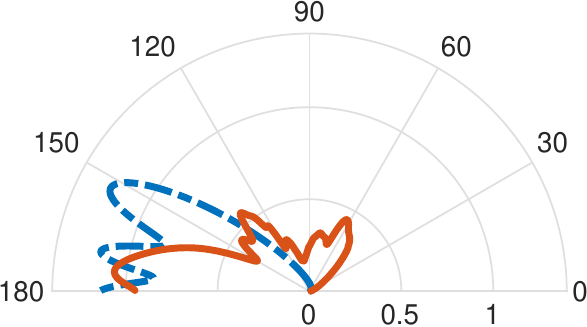}
    \caption{$y_2 = 0$, $k=50$}
    \end{subfigure}
    \begin{subfigure}{0.47\textwidth}
    \centering
    \includegraphics[width=\linewidth]{FarFieldDirectivitytheta0p25y0p25k10h2M0.pdf}
    \caption{$y_2 = 0.25$, $k=10$}
    \end{subfigure}
    \hfill
    \begin{subfigure}{0.47\textwidth}
    \centering
    \includegraphics[width=\linewidth]{FarFieldDirectivitytheta0p25y0p25k50h2M0.pdf}
    \caption{$y_2 = 0.25$, $=50$}
    \end{subfigure}
    \begin{subfigure}{0.47\textwidth}
    \centering
    \includegraphics[width=\linewidth]{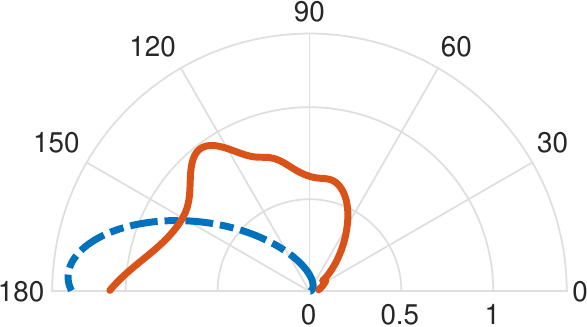}
    \caption{$y_2=0.75$, $k = 50$}
    \end{subfigure}
    \hfill
    \begin{subfigure}{0.47\textwidth}
    \centering
    \includegraphics[width=\linewidth]{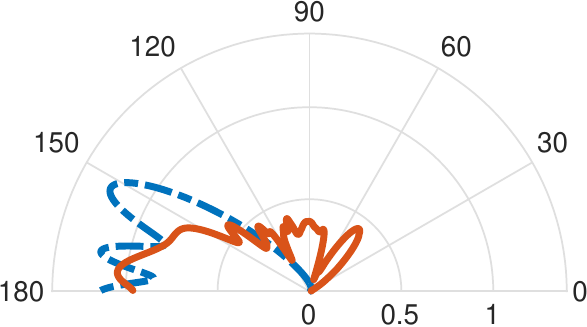}
    \caption{$y_2 = 0.75$, $k=50$}
    \end{subfigure}
    \caption{Far-field directivity patterns of the scattered pressure due to
    the baseline and serrated edge of $h=2$. The source is located at $r=1$ and
    $\theta = \pi/4$ in the $y_2=0$, $y_2=0.25$ and $y_2=0.75$ plane,
    respectively.}
    \label{fig:SpanwiseEffect}
\end{figure}

\add{Both figures~\ref{fig:AmplitudeEffects} and \ref{fig:AmplitudeEffects2}
show that when serrations are used evident noise increase occurs at some
observer angles, which is particularly pronounced at high frequencies. This may
seem somewhat surprising as serrations are in fact used to reduce rather than
increase TE noise. This apparent contradiction arises because the Green's
function developed in this paper is for a simple acoustic point source, whereas
in practical applications the sources are of a distribution type and
characterized by hydrodynamic length scales that are much shorter than the
acoustic wavelengths. It is known that the destructive interference introduced
by serrations, which hinges on the fact that the hydrodynamic fluctuations are
characterized by short length scales, plays a significant role in reducing TE
noise~\citep{Howe1991a,Lyu2016a,Jaworksi2020}. Once these hydrodynamic length
scales are taken into account, significant noise reduction can be expected.
Therefore, the noise directivity observed in this paper is not to be confused
with that of the total TE noise.}

\subsection{Effects of the source position}
Figures~\ref{fig:AmplitudeEffects} and \ref{fig:AmplitudeEffects2} show the
scattered pressure directivity for sound sources located in the serration tip
plane. It would be interesting to understand how the scattering characteristics
change when the spanwise location of the sound source changes.
Figure~\ref{fig:SpanwiseEffect} shows the directivity patterns for sound
sources at $\theta=\pi/4$ but at different spanwise locations when $M=0$.
Figure~\ref{fig:SpanwiseEffect}(a-b) shows the noise directivity for sources
located in $y_2 = 0$. We see that noise is only slightly increased at low
observer angles, and similarly slightly reduced at high angles. This is because
the sound sources are located off the serration tip plate, therefore the
additional (weakened) reflection due to the extended (removed) surface is
weaker. Significant noise increase occurs as the source moves to the tip plane,
as explained in figure~\ref{fig:AmplitudeEffects}. When the spanwise location
moves to $0.5$, we expect the resulting directivity to be identical to that at
$y_2 = 0$ due to symmetry. When the source position moves to $y_2 = 0.75$, i.e.
in the serration root plane, the resulting directivity is shown in
figure~\ref{fig:SpanwiseEffect}(e-f). It seems surprising to see an even
broader lobe in the range of $45^\circ < \Theta < 135^\circ$ as the source is
not directly above any rigid surface. However, we note that compare to the
baseline trailing edge, there do exist additional rigid surfaces downstream,
albeit slightly off the plane. Besides, constructive interference may be
expected as the two serration teeth are geometrically symmetric with respect to
$y_2=0.75$. If this were true, we would expect a less pronounced noise increase
at high frequencies, because the scattering will become localized and the
scattered pressure drops more rapidly as the frequency increases. This is
indeed the case, as shown in figure~\ref{fig:SpanwiseEffect}(f).

 \begin{figure}
    \centering
    \begin{subfigure}{0.47\textwidth}
    \centering
    \includegraphics[width=\linewidth]{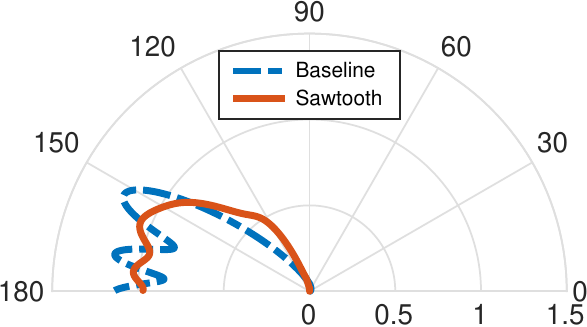}
    \caption{$r=5, k=10$}
    \end{subfigure}
    \hfill
    \begin{subfigure}{0.47\textwidth}
    \centering
    \includegraphics[width=\linewidth]{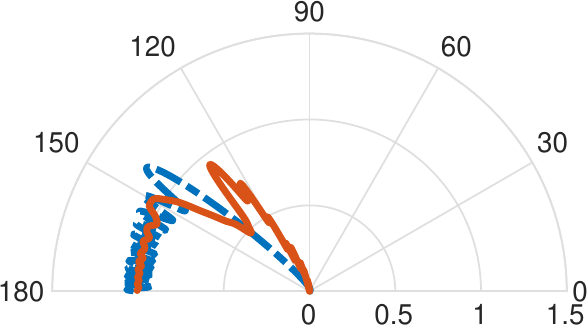}
    \caption{$r=5, k=50$}
    \end{subfigure}
    \begin{subfigure}{0.47\textwidth}
    \centering
    \includegraphics[width=\linewidth]{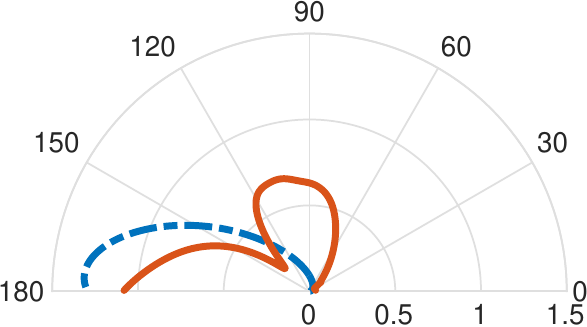}
    \caption{$r=1, k=10$}
    \end{subfigure}
    \hfill
    \begin{subfigure}{0.47\textwidth}
    \centering
    \includegraphics[width=\linewidth]{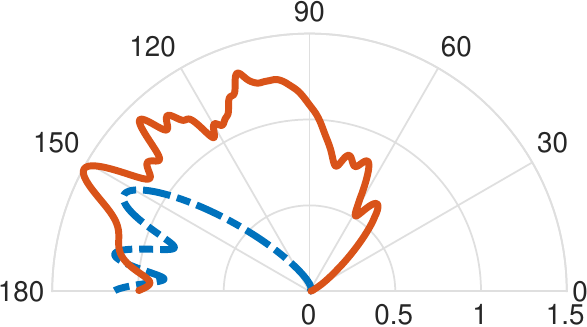}
    \caption{$r=1, k=50$}
    \end{subfigure}
    \begin{subfigure}{0.47\textwidth}
    \centering
    \includegraphics[width=\linewidth]{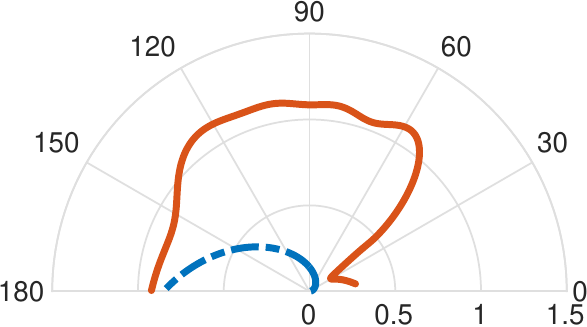}
    \caption{$r=0.2, k = 10$}
    \end{subfigure}
    \hfill
    \begin{subfigure}{0.47\textwidth}
    \centering
    \includegraphics[width=\linewidth]{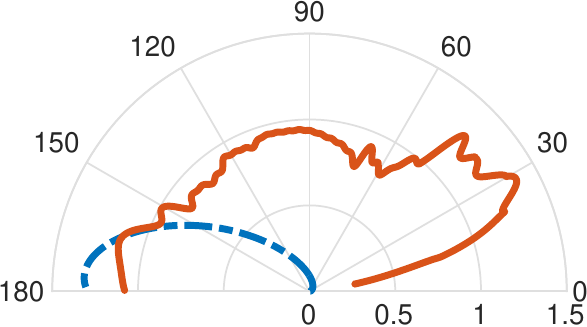}
    \caption{$r=0.2, k = 50$}
    \end{subfigure}
    \caption{Far-field directivity patterns of the scattered pressure due to
    the baseline and serrated edge ($h=2$). The source is located at $\theta =
    \pi/4$ in the $y_2=0.25$ plane (tip plane) and $r$ takes the value of $5$,
    $1$ and $0.2$, respectively.}
    \label{fig:DistanceEffects}
\end{figure}

Apart from varying the spanwise location of the source, we are also interested
in the effects of varying the radial distance in the $(y_1, y_3)$ plane, i.e.
$\sqrt{y_1^2+y_3^2}$. As we consider the case of $M=0$, this distance is the
same as $r$. For baseline TE scattering, the only characteristic length scale
is the sound wavelength (apart from $r$). Therefore, the scattering can only
depend on the non-dimensional number $kr$. In other words, decreasing $r$ would
be equivalent to increasing the frequency. The use of serrated trailing edges,
however, introduces two additional lengths scales, i.e. the wavelength and
amplitude of the serration. Therefore, we expect a change in the scattering
characteristics as $r$ varies. Figure~\ref{fig:DistanceEffects} shows the
directivity patterns for a sound source located at $\theta=\pi/4 $ but various
$r$ in the tip plane. In particular, figure~\ref{fig:DistanceEffects}(a-b)
shows the scattered directivity when the source is far from the edge at $r=5$.
We see that the difference between the straight and serrated cases is not
pronounced. This is expected because the source is relatively far away from the
edge, and therefore the enhanced and weakened reflection due to the extended or
removed surface would not be strong. Considerable change, however, occurs when
the source moves closer to the edge to $r=1$, as can be seen from
figure~\ref{fig:DistanceEffects}(c-d). At this distance, the source effectively
sees parts of the reflection surface extended while others removed, and
consequently the noise radiation is amplified or reduced at low or large
observer angles. It is interesting to see that the baseline directivity
patterns in figures~\ref{fig:DistanceEffects}(a) and
\ref{fig:DistanceEffects}(d) are identical. Indeed, without the additional
length scales introduced by the serrations, the scattering would only depend on
$kr$, which attains equal values in both cases. However, when serrations are
used, we see that the directivity patterns are no longer the same even though
$kr$ remains identical. In particular, a closer source leads to more pronounced
modifications to the directivity patterns. When the source continues to move to
the close vicinity of the edge, we would expect even more pronounced effects
induced by the serrations. Indeed, comparing
figures~\ref{fig:DistanceEffects}(c-d) and \ref{fig:DistanceEffects}(e-f), we
see that the low-angle amplification is substantial. Considering in the
serrated case the point source is immediately above a rigid plate that does not
exist in the baseline case, such strong modifications to the low-angle
directivity can be expected.

\subsection{Effects of the Mach number}
\begin{figure}
    \centering
    \begin{subfigure}{0.47\textwidth}
    \centering
    \includegraphics[width=\linewidth]{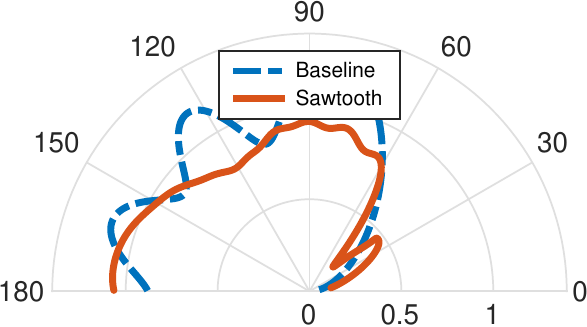}
    \caption{$M=0.2, k=10$}
    \end{subfigure}
    \hfill
    \begin{subfigure}{0.47\textwidth}
    \centering
    \includegraphics[width=\linewidth]{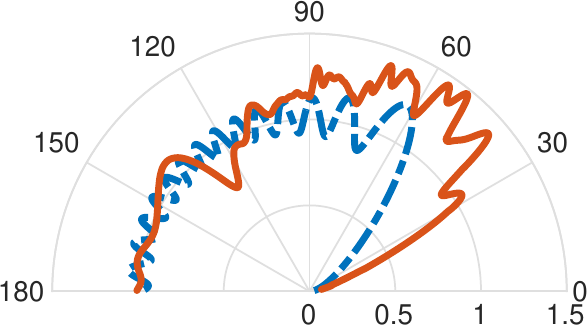}
    \caption{$M=0.2, k=50$}
    \end{subfigure}
    \begin{subfigure}{0.47\textwidth}
    \centering
    \includegraphics[width=\linewidth]{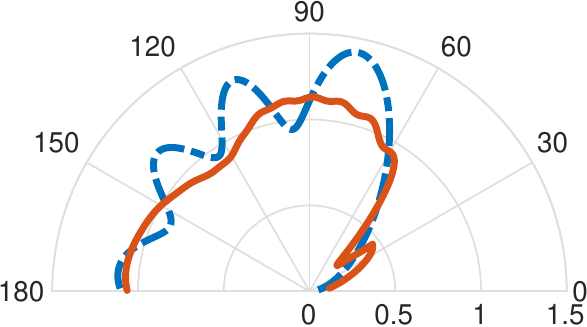}
    \caption{$M=0.5, k=10$}
    \end{subfigure}
    \hfill
    \begin{subfigure}{0.47\textwidth}
    \centering
    \includegraphics[width=\linewidth]{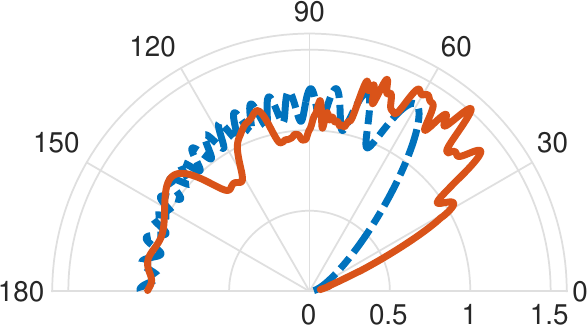}
    \caption{$M=0.5, k=50$}
    \end{subfigure}
    \begin{subfigure}{0.47\textwidth}
    \centering
    \includegraphics[width=\linewidth]{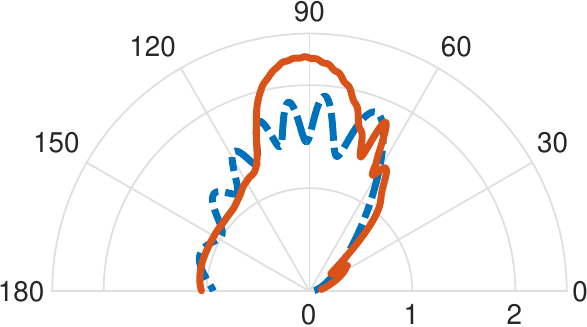}
    \caption{$M=0.8, k=10$}
    \end{subfigure}
    \hfill
    \begin{subfigure}{0.47\textwidth}
    \centering
    \includegraphics[width=\linewidth]{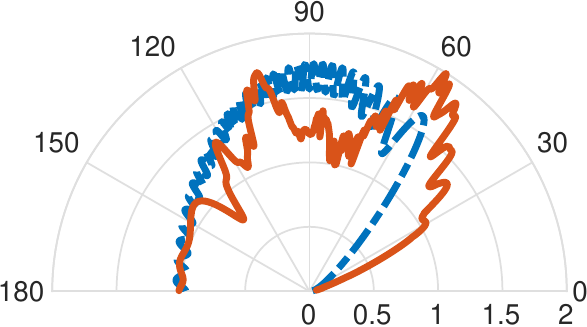}
    \caption{$M=0.8, k=50$}
    \end{subfigure}
\caption{Far-field directivity patterns of the scattered pressure due to the
baseline and serrated edges at various Mach numbers. The serration has an
amplitude of $h=2$. The source is located at $r=1$ and $\theta = 3\pi/4$ in the
$y_2=0.25$ plane.}
\label{fig:MachEffects}
\end{figure}

The effect of varying Mach numbers can be similarly studied.
Figure~\ref{fig:MachEffects} shows the far-field directivity patterns for a
sound source located at $(\cos(3\pi/4), 0.25, \sin(3\pi/4))$ but with various
Mach numbers. The serration has an amplitude of $h=2$.
Figure~\ref{fig:MachEffects}(a-b) shows the directivity at $M=0.2$. Compared
with figure~\ref{fig:AmplitudeEffects2}(c-d), we see that little change occurs
at such a Mach number in both straight and serrated cases. When we increase the
Mach number to $M=0.5$, we start to see that the mean flow tends to increase
the radiation magnitude at side angles (around $\Theta=90^\circ$), whereas no
amplification or reduction seems to occur at $\Theta=0^\circ$ and
$\Theta=180^\circ$ at all. The consequence is that the directivity pattern is
effectively squeezed into a thinner lobe. Such a tendency is consistent between
the baseline and serrated cases, and is much more evident at the Mach number of
$0.8$, which is shown in figure~\ref{fig:MachEffects}(e-f). For example, we see
that the baseline directivity peaks at around $\Theta=90^\circ$ and obtains a
value of $1.7$, while the pressure magnitude at $\Theta=180^\circ$ remains to
be $1$. 
%Such effects are also predicted in earlier works~\citep{Lyu2016a}. 

\begin{figure}
    \centering
    \includegraphics[width=0.7\textwidth]{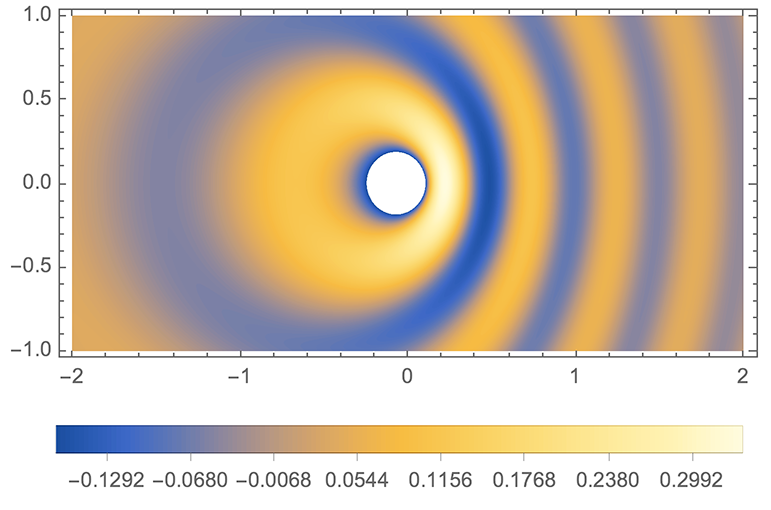}
    \caption{Instantaneous pressure distribution due to a moving source
    travelling from left to right at a Mach number of $0.5$. The instantaneous
    position of the source is shown by the white circle.}
    \label{fig:Doppler}
\end{figure}

The fact that convection amplification occurs at side angles instead of forward
angles appears very strange and even contradictory to the well-known Doppler
effects. It is well established that sound is amplified in the forward
($\Theta<\pi/2$) arc and reduced in the backward ($\Theta > \pi/2$) arc by a
factor of $(1-M\cos\Theta)^{-1}$ when a point source is in motion. This would
also imply that no sound amplification occurs at $\Theta=90^\circ$. To
understand why such a discrepancy occurs, we start with a simple case that a
point source travels from {\em left} to {\em right} with a uniform Mach number
$M$. We can choose the coordinate frame to be static relative to the medium or
to the point source. The former is often used in the classical Doppler analysis
while the latter in the TE noise modelling. The instantaneous pressure field
generated by such a moving source can be calculated analytically and plotted in
figure~\ref{fig:Doppler}, where the Mach number is chosen to be $M=0.5$.
\add{Note that figure~\ref{fig:Doppler} is a standard result and is only
included here for the following illustration purposes.}

In the coordinate frame that remains still to the source (e.g. with its origin
fixed on the point source), the acoustic pressure at a fixed distance to the
source does not possess the same phase (see figure~\ref{fig:Doppler}). However,
although not shown here in detail, it can be verified that the magnitude of the
pressure is amplified by $\beta^{-1}$ at $\Theta=90^\circ$ and remains
unchanged at $\Theta=0^\circ$ and $180^\circ$. This is precisely what we
observe for our Green's function, where the TE noise source effectively remains
still at the origin. In fact, the maximal value obtained at $\Theta=90^\circ$
in figure~\ref{fig:MachEffects}(e) is around $1.7$ at $M=0.8$, and this is
consistent with the amplification ratio of $\beta^{-1}=1.67$. However, in the
coordinate frame that remains still relative to the medium, we are often
interested in measuring the sound at a fixed distance to its emitting position.
Clearly they must have the same phase, as shown for example by the dark circles
in figure~\ref{fig:Doppler}. Figure~\ref{fig:Doppler} clearly shows that the
emitted sound is amplified in the forward arc and reduced in the backward arc.
Although not shown here in detail, it can be readily verified that the
amplification ratio shown in figure~\ref{fig:Doppler} is precisely
$(1-M\cos\Theta)^{-1}$.

In summary, the seemingly strange convective behaviour shown in
figure~\ref{fig:MachEffects} is consistent with the classical Doppler effects,
and the apparent contradiction is due to the use of a different coordinate
frame. This shows that proper corrections have to be applied when wind tunnel
results are used to predict the noise radiation directivity of practical
applications.

\section{Conclusion}
In this paper, we develop an analytical Green’s function for the acoustic
scattering by serrated edges, which is in a closed form and is applicable to
both TE and LE scatterings. The standard adjoint Green's function technique is
used to formulate the derivation into a pressure scattering problem under a
plane wave incidence. The total pressure field is then decomposed into an
incident, a hypothetically reflected and a reflection-removed scattered part.
The scattered field is subsequently solved using the Wiener-Hopf method. It is
shown that a recent kernel factorization used in TE and LE noise modelling
appears not uniformly valid, but may be used as a reasonable approximation in
the high frequency regime. Closed-form analytical Green's functions in the form
of Fresnel integrals are obtained for any arbitrary piecewise linear
serrations. Numerical integration is used to validate the derivations, which
shows excellent agreement with the results given by the analytical formula. The
Green's function is then compared with the scattered pressure calculated using
FEM in COMSOL. The results clearly demonstrate that the $E_n(s)$ assumption is
problematic at small $kh$ values, but serves as a reasonably good
approximations at large values. The noise directivity patterns are studied as a
function of the frequency, serration amplitude, source position and Mach
number, respectively. It is shown that the use of serrations enhances noise
radiation at low observer angles. The strength of this enhancement increases as
the frequency increases. On the other hand, slight noise weakening may occur at
large observer angles, but such effects diminish as the frequency increases.
These directivity changes may be understood from the perspective of an extended
or removed rigid reflection surface, and are therefore more evident when the
source moves closer to the edge. Increasing the Mach number appears to amplify
sound at side angles but not at $\Theta=0^\circ$ and $\Theta=90^\circ$. This
seemingly strange behaviour is the consequence of using a coordinate frame that
is static relative to the TE noise source. 

Due to its analytical nature, the Green's function can be evaluated quickly.
Because of symmetry, it is applicable to both leading- and trailing-edge
scatterings. Such a Green's function would be particularly suitable for
developing a leading- or trailing-edge noise model that is both highly
efficient and three-dimensionally accurate. More importantly, with a proper
knowledge of the turbulence statistics inside a boundary layer, the Green's
function may be used to consider the effects of non-frozen turbulence on TE
noise and its reduction using serrations. These form part of our future work.
%Upon dividing both sides of (\ref{equ:WienerHopfEquation}) by
%$\sqrt{s+\kappa_n}$ and cancelling out the pole in the second term of
%(\ref{equ:WienerHopfEquation}), we have \begin{equation} \begin{aligned}
%\sqrt{s - \kappa_n} \mathcal{G}^-_n(s) + \frac{k_3 E_n(s)}{(s-k_1)}
%\frac{1}{\sqrt{k_1 + \kappa_n}} & = - \frac{\mathcal{G}_n^\prime(s)}{\sqrt{s +
%\kappa_n}} - \frac{k_3 E_n(s)}{(s-k_1)}  \left(\frac{1}{\sqrt{s + \kappa_n}} -
%\frac{1}{\sqrt{k_1 + \kappa_n}}\right) \\ & \equiv E(s). \end{aligned}
%\label{equ:Match} \end{equation} It is easy to verify that $E(s) \equiv 0$ by
%invoking the Louville Theorem, and we are left with, 
\section*{Acknowledgement}
The first author (BL) wishes to gratefully acknowledge the financial support
from Laoshan Laboratory under the grant number of LSKJ202202000.

\textbf{Declaration of Interests.} The authors report no conflict of interest.

\appendix
\section{Analytical evaluation of $H_n^\pm(r_i, \theta_i)$ ($n\ne 0$)}
\label{app:HnEvaluation}
In order to evaluate (\ref{equ:ThetaIntegration}) i.e.
\begin{equation}
    D_n^\pm(r_i, \theta_i) 
    = -\sqrt{\frac{2}{\kappa_n}}  
    \int_{-\infty}^{\infty}
    \frac{\sin\frac{1}{2}(\theta_i + \i t)}
    {\cos(\theta_i+\i t) + \cos\Theta_n^{\pm}}
    \e^{\i \kappa_n r_i \cosh t} \ud t,
    \label{equ:appendixInt}
\end{equation}
we note that 
\begin{IEEEeqnarray}{rCl}
    \label{equ:trigonometricIdentities}
    \cos(\theta_i+\i t) + \cos\Theta_n^{\pm} 
    &=& 2 \cos \frac{1}{2}(\theta_i+\i t+\Theta_n^\pm)
    \cos \frac{1}{2}(\theta_i+\i t-\Theta_n^\pm),
    \IEEEyesnumber\IEEEyessubnumber\\
    2\sin \frac{1}{2}(\theta_i + \i t) \sin \frac{1}{2}\Theta_n^\pm 
    &=& \cos \frac{1}{2}(\theta_i + \i t -\Theta_n^\pm) 
    - \cos \frac{1}{2}(\theta_i + \i t +\Theta_n^\pm).
    \IEEEyessubnumber
\end{IEEEeqnarray}
Making use of (\ref{equ:trigonometricIdentities}), the integrand in
(\ref{equ:appendixInt}) can be written as
\begin{IEEEeqnarray}{rl}
    \frac{1}{4 \sin \frac{1}{2}\Theta_n^\pm}
    \left[
	\frac{1} {\cos \frac{1}{2}(\i t+\theta_i+\Theta_n^\pm)}
	- \frac{1}{\cos \frac{1}{2}(\i t + \theta_i-\Theta_n^\pm)}
    \right]
\end{IEEEeqnarray}
and consequently
\begin{IEEEeqnarray}{rl}
    D_n^\pm(r_i, \theta_i) 
    = -\sqrt{\frac{2}{\kappa_n}}  
    \frac{1}{\sin \frac{1}{2}\Theta_n^\pm}
    \Big[N(\theta_i + \Theta_n^\pm) - N(\theta_i - \Theta_n^\pm)\Big],
    \label{equ:fractionedInt}
\end{IEEEeqnarray}
where
\begin{equation}
    N(\psi) = \int_{-\infty}^{\infty} 
    \frac{1} {4 \cos \frac{1}{2}(\i t + \psi )}
    \e^{\i \kappa_n r_i \cosh t} \ud t.
    \label{equ:NIntegral}
\end{equation}

Equation~(\ref{equ:NIntegral}) can be evaluated by multiplying both the
numerator and denominator of the integrand by $\cos \frac{1}{2} (\i t -
\psi)$ and making use of the odd and even properties of the integrand,
resulting in
\begin{equation}
    N(\psi) = \int_0^\infty
    \frac{\cosh \frac{1}{2}t \cos \frac{1}{2}\psi} 
    {\cosh t + \cos\psi }
\e^{\i \kappa_n r_i \cosh t} 
     \ud t.
\end{equation}
Letting $\tau = \sinh \frac{1}{2}t$ and $\cosh t = 2\tau^2 + 1$, one can show
that~\citep{Noble1958}
\begin{equation}
    N(\psi) = 
    \e^{-\i \kappa_n r_i \cos\psi}
    \cos \frac{1}{2}\psi 
    \int_0^\infty
    \frac{\e^{\i 2\kappa_n r_i (\tau^2 + \cos^2 \frac{1}{2}\psi)}}
    {\tau^2 + \cos^2 \frac{1}{2}\psi}
    \ud \tau.
    \label{equ:N}
\end{equation}
It is well known that 
\begin{equation}
    \int_0^\infty
    \e^{\i 2\kappa_n r_i (2\tau^2 + \cos^2 \frac{1}{2}\psi)}
    \ud \tau 
    =
    \frac{1}{2} 
    \e^{\i 2 \kappa_n r_i \cos^2 \frac{1}{2}\psi}
    \sqrt{\frac{\pi}{2\kappa_n r_i}} 
    \e^{-\i \frac{\pi}{4}}.
    \label{equ:StandardFresnelInt}
\end{equation}
Integrating both side of (\ref{equ:StandardFresnelInt}) with respect to
$r_i$, one can show that (\ref{equ:N}) can be evaluated to
\begin{equation}
    N(\psi) = \e^{-\i\kappa_n r_i \cos\psi}
    \sqrt{\pi} 
    \e^{-\i \frac{\pi}{4}}
    F(\sqrt{2\kappa_n r_i} \cos \frac{1}{2}\psi),
    \label{equ:Nresult}
\end{equation}
where the Fresnel integral $F$ is defined by
(\ref{equ:FresnelIntegral}). Substituting (\ref{equ:Nresult})
into (\ref{equ:fractionedInt}), we obtain the final solution
\begin{equation}
    D_n^\pm(r_i, \theta_i) 
    = -\pi\sqrt{\frac{2}{\kappa_n}}  
     \frac{I(\kappa_n r_i, \theta_i, \Theta_n^\pm)}{\sin
    \frac{1}{2}\Theta_n^\pm},
\end{equation}
where function $I(\kappa_n r_i, \theta_i, \Theta_n^\pm)$ is defined in
(\ref{equ:standardIn}). \add{Using the above integral, we can quickly obtain
(\ref{equ:Hfun}).}

\section{Analytical evaluation of $H_0(r_i, \theta_i)$}
\label{app:H0Integration}
When $n=0$, (\ref{equ:Hn}) reduces to 
\begin{equation}
    H_0(r_i, \theta_i) 
    =-\frac{\i}{2\overline{h}}
    \int_{-\infty}^{\infty}
    \frac{1}{(s-k_1)^2}
    \frac{1}{\sqrt{s - \kappa_0}}
    \e^{(-i s \cos \theta_i  - \gamma_0  \sin\theta_i)r_i} \ud s.
    \label{equ:H0}
\end{equation}
Upon deforming onto $P_2$, one obtains
\begin{equation}
    H_0(r_i, \theta_i) 
    = -\frac{\i}{\sqrt{2\kappa_0} \kappa_0 \overline{h}}
    \int_{-\infty}^{\infty}
    \frac{\sin\frac{1}{2}(\theta_i + \i t)}
    {\left(\cos(\theta_i+\i t) + \cos\Theta_0\right)^2}
    \e^{\i \kappa_0 r_i \cosh t} \ud t.
	\label{equ:H0ThetaIntegration}
\end{equation}
In order to evaluate (\ref{equ:H0ThetaIntegration}), we make use of the
same trigonometric identity shown in (\ref{equ:trigonometricIdentities})
such at
\begin{equation}
    H_0(r_i, \theta_i) 
    = -\frac{\i}{\sqrt{2\kappa_0} \kappa_0 \overline{h}}
    \frac{1}{\sin \frac{1}{2}\Theta_0}
    \Big[
	M\left(\theta_i + \Theta_0, \theta_i - \Theta_0\right)
	-M\left(\theta_i - \Theta_0, \theta_i + \Theta_0\right)
    \Big],
    \label{equ:H0InTermsOfM}
\end{equation}
where
\begin{equation}
    M(\psi_1, \psi_2) = 
    \int_{-\infty}^{\infty}
    \frac{1}{8\cos^2 \frac{1}{2}(\i t + \psi_1)\cos \frac{1}{2}(\i t + \psi_2)}
    \e^{\i \kappa_0 r_i \cosh t} \ud t.
    \label{equ:M}
\end{equation}

To evaluate (\ref{equ:M}), we multiply both the numerator and
denominator by $\cos^2 \frac{1}{2}(\i t - \psi_1)\cos \frac{1}{2}(\i t -
\psi_2)$, expand using trigonometric identities and make use of the odd and
even properties of the integrand to rewrite (\ref{equ:M}) as
\begin{equation}
    \int_0^\infty 
    \frac{(1+\cosh t \cos\psi_1)(\cosh \frac{1}{2}t + \cos \frac{1}{2}\psi_2
    - \sinh t \sin\psi_1 \sinh \frac{1}{2}t \sin \frac{1}{2}\psi_2)}
    {(\cosh t + \cos\psi_1)^2 (\cosh t + \cos\psi_2)}
    \e^{\i \kappa_0 r_i \cosh t} \ud t.
    \label{equ:Mexpanded}
\end{equation}
Let $\tau=\sinh \frac{1}{2}t$ and hence $\cosh t = 1 + 2\tau^2$,
(\ref{equ:Mexpanded}) reduces to
\begin{equation}
    M(\psi_1, \psi_2)  
    =
    \int_0^\infty
    \frac{\cos(\psi_1 + \frac{\psi_2}{2}) \tau^2 
    + \cos^2 \psi_1 \cos \frac{\psi_2}{2}}
    {2(\tau^2 + \cos^2 \frac{\psi_1}{2})^2 (\tau^2 + \cos^2 \frac{\psi_2}{2})}
    \e^{\i \kappa_0 r_i (2\tau^2 + 1)} \ud t.
    \label{equ:MInTau}
\end{equation}
Note that the rational expression in the integrand of (\ref{equ:MInTau})
can be expanded in terms of partial fractions, i.e.
\begin{equation}
    \frac{\cos(\psi_1 + \frac{\psi_2}{2}) \tau^2 
    + \cos^2 \psi_1 \cos \frac{\psi_2}{2}}
    {2(\tau^2 + \cos^2 \frac{\psi_1}{2})^2 (\tau^2 + \cos^2 \frac{\psi_2}{2})}
    = \frac{A_1}{\tau^2 + \cos^2 \frac{1}{2}\psi_1}
    +\frac{A_2}{(\tau^2 + \cos^2 \frac{1}{2}\psi_1)^2}
    + \frac{A_3}{\tau^2 + \cos^2 \frac{1}{2}\psi_2},
\end{equation}
where 
\begin{IEEEeqnarray}{rCl}
    \label{equ:paritalFraction}
    A_1 
    &=& 
    \frac{-\cos \frac{1}{2}\psi_2}
    {2\sin^2 \frac{1}{2}(\psi_1 - \psi_2)}, \IEEEyesnumber\IEEEyessubnumber\\
    A_2 
    &=& \frac{\sin \frac{1}{2}\psi_1 \cos^2 \frac{1}{2}\psi_1}
    {\sin \frac{1}{2}(\psi_1 - \psi_2)}, \IEEEyessubnumber\\
    A_3 
    &=& -A_1. \IEEEyessubnumber
\end{IEEEeqnarray}
The first and third partial fractions are similar to (\ref{equ:N}),
therefore can be readily evaluated.  The second term can be evaluated by
integrating (\ref{equ:StandardFresnelInt}) twice with respect to $r_i$,
so that the $M(\psi_1, \psi_2)$ reduces to,
\begin{IEEEeqnarray}{rCl}
    M(\psi_1, \psi_2) 
    &=&  A_1 
    \e^{-\i \frac{\pi}{4}}
    \frac{\sqrt{\pi}}{\cos \frac{1}{2} \psi_1}
    \e^{-\i \kappa_0 r_i \cos\psi_1} 
    F(\sqrt{2\kappa_0 r_i} \cos \frac{1}{2}\psi_1)
    \nonumber\\
    && \negmedspace {} +  A_2 
    \e^{-\i \frac{\pi}{4}}
    \frac{\sqrt{\pi}}{\cos^3 \frac{1}{2} \psi_1}
    \e^{-\i \kappa_0 r_i \cos\psi_1} 
    \Big(
	2\i\kappa_0 r_i \cos^2 \frac{1}{2}\psi_1 
	F(\sqrt{2\kappa_0 r_i} \cos \frac{1}{2}\psi_1) \nonumber\\
    &&
    \quad \quad \quad \quad \quad \quad
    + 
	\frac{1}{2} \sqrt{2\kappa_0 r_i \cos^2 \frac{1}{2}\psi_1} 
	\e^{\i 2\kappa_0 r_i \cos^2 \frac{1}{2}\psi_1} 
	+
	\frac{1}{2}F(\sqrt{2\kappa_0 r_i} \cos \frac{1}{2}\psi_1)
    \Big)\nonumber\\
    && \negmedspace {} +  A_3 
    \e^{-\i \frac{\pi}{4}}
    \frac{\sqrt{\pi}}{\cos \frac{1}{2} \psi_2}
    \e^{-\i \kappa_0 r_i \cos\psi_2} 
    F(\sqrt{2\kappa_0 r_i} \cos \frac{1}{2}\psi_2). \IEEEeqnarraynumspace
    \label{equ:FinalM}
\end{IEEEeqnarray}

Substituting (\ref{equ:FinalM}) into (\ref{equ:H0InTermsOfM}) and
collecting common terms, we show that $H_0(r_i, \theta_i)$ can be calculated
explicitly as 
\begin{IEEEeqnarray}{rl}
    H_0(r_i, \theta_i) 
    =  \frac{\i\pi}
    {4\sqrt{2\kappa_0}\kappa_0 \bar{h} }  
    \frac{1}{\sin^2\frac{\Theta_0}{2}}
    \left[
	\frac{I(\kappa_0 r_i, \theta_i;
	\Theta_0)}{\sin\frac{\Theta_0}{2}}\right.
	&- \frac{(2\i\kappa_0 r_i) J(\kappa_0 r_i, \theta_i;\Theta_0)}
	{\cos\frac{\Theta_0}{2}} \nonumber\\
    & \left. -\frac{\e^{-\i \frac{\pi}{4}}}{\sqrt{\pi} }
	2\sqrt{2\kappa_0 r_i} \sin\frac{\theta_i}{2} \e^{\i \kappa_0 r_i}
    \right],
     \IEEEeqnarraynumspace
\end{IEEEeqnarray}
where the subscript $i$ takes the value of either $t$ or $r$. The Fresnel
function $I(\kappa_0 r_i, \theta_i; \Theta_0)$ is defined in
section~\ref{sec:GreensFunction}, and $J(\kappa_0 r_i, \theta_i; \Theta_0)$ is
very similar to $I(\kappa_0 r_i, \theta_i; \Theta_0)$, i.e.
\begin{IEEEeqnarray}{rCl}
    J(\kappa_0 r_i, \theta_i; \Theta_0) = 
    \frac{1}{\sqrt{\pi}} \e^{-\i \frac{\pi}{4}} 
     &&\left[
	 \sin(\Theta_0+\theta_i)
	 \e^{-\i \kappa_0 r_i \cos(\Theta+\theta)} 
     F(\sqrt{2\kappa_0 r_i} \cos \frac{\Theta_0+\theta_i}{2})\right.
     \nonumber\\
     &&\left.-\sin(\Theta_0-\theta_i)
	 \e^{-\i \kappa_0 r_i \cos(\Theta_0-\theta_i)} 
     F(\sqrt{2\kappa_0 r_i} \cos \frac{\Theta_0-\theta_i}{2}) \right].
     \IEEEeqnarraynumspace
\end{IEEEeqnarray}

\section{Green's function for other piecewise linear serration profiles}
\label{app:OtherGs}
As mentioned in section~\ref{sec:GreensFunction}, the Green's function can be
calculated analytically for arbitrary piecewise linear serrations. For other
serration profiles, the scattered Green's function $G_s$ can still be written
as
\begin{IEEEeqnarray}{rCl}
    G_s(r,\theta, y_2) = \frac{1}{2\pi}\e^{\i k M y_1 / \beta^2}
    \sum_{n = -\infty}^{\infty} &&
    -\i(\sqrt{k_1 - \kappa_n}) \nonumber\\
		       && \times \left[
			   \int_{-\infty}^{\infty}
			   \frac{E_n(s)}{s-k_1}  
			   \frac{1}{\sqrt{s - \kappa_n}}
			   \e^{(-i s  \cos \theta  - \gamma_n  \sin\theta) r} \ud s  
		       \right], \IEEEeqnarraynumspace
\end{IEEEeqnarray}
where the integral path is given by $P_0$ as shown in
figure~\ref{fig:integralPath} and $E_n(s)$ is defined by
(\ref{equ:EnDefinition}). For square-shaped serrations 
\begin{equation}
    E_n(s) = \frac{2\sin \frac{1}{2}n\pi}{n\pi} 
    \e^{-\i \frac{1}{2}n\pi}
    \cos[(s-k_1)\overline{h} + \frac{n\pi}{2}].
\end{equation}
Note that $E_n(s)$ can be readily written as
\begin{equation}
    E_n(s) = 
    \frac{\sin \frac{1}{2}n\pi}{n\pi} 
    \e^{\i \frac{1}{2}n\pi}
    \left( 
	\e^{\i [(s-k_1)\overline{h} - \frac{n\pi}{2}]}
	+\e^{-\i [(s-k_1)\overline{h} - \frac{n\pi}{2}]}
    \right). 
    \label{equ:squaredEn}
\end{equation}
\add{Equation~(\ref{equ:squaredEn}) is similar to (\ref{equ:EnSawtooth}) (after
trigonometric expansions), and the Green's function can be calculated in a
similar manner (without performing additional integration) to yield}
\begin{IEEEeqnarray}{rCl}
    G_s(r, \theta, y) &=& \frac{1}{2\pi}\e^{\i k M y_1 / \beta^2}\sum_{n = -\infty}^{\infty}
    -\i(\sqrt{k_1 - \kappa_n})
    \e^{\i \chi_n y_2}
    \frac{\sin \frac{1}{2}n\pi}{n\pi} 
    \e^{\i \frac{1}{2} n\pi}
    \nonumber\\
		      && \times
		      \left( \exp\left( -\i(k_1
			  \bar{h}+\frac{n\pi}{2})\right) D_0(r_t, \theta_t) 
			  - \exp\left( \i(k_1
		      \bar{h}+\frac{n\pi}{2})\right)D_0(r_r, \theta_r)\right),
		      \IEEEeqnarraynumspace 
\end{IEEEeqnarray}
where $D_0(r_i, \theta_i) \equiv D_0^+(r_i, \theta_i)\equiv D_0^-(r_i,
\theta_i)$ is defined in (\ref{equ:Hpm}). Note that $D_0(r_i, \theta_i)$ does
not depend on $n$, which is different from that for sawtooth serrations.
\add{In fact, we can show that $E_n(s)$ for arbitrary piecewise linear
serrations can be written in a similar form as those shown in
(\ref{equ:squaredEn}),} therefore their corresponding Green's functions can be
calculated similarly in a straightforward manner. We omit a repetitive
description here.

\section{An alternative decomposition}
\label{app:AnotherDecomposition}
When we use the decomposition $G^a = p_{in} + G_s$ instead, the scattered wave
$G_s(\boldsymbol{y};\boldsymbol{x}, \omega)$ satisfies
\begin{equation}
    \beta^2 \frac{\partial^2 G_s}{\partial y_1^2} 
    + \frac{\partial^2 G_s}{\partial y_2^2} + \frac{\partial^2 G_s}{\partial y_3^2} - 2\i k M \frac{\partial G_s}{\partial y_1} 
    + k^2 G_s = 0,
    \label{equ:WaveEquation}
\end{equation}
and the following boundary conditions, 
\begin{IEEEeqnarray}{rCl}
	\left.\frac{\partial G_s}{\partial y_3}\right|_{y_3=0} 
	&=&  i k_3 \e^{-\i k_1 y_1 / \beta} \e^{\i\frac{kM}{\beta^2}y_1} 
	\e^{-\i k_2 y_2 },\quad y_1 < hF(y_2);\IEEEyesnumber\IEEEyessubnumber \\
	 G_s|_{y_3 = 0} 
	&=& 0, \quad y_1 > hF(y_2); \IEEEyessubnumber\\
	 G_s|_{y_2 = 0} 
	&=& G_s|_{y_2 = 1} \e^{\i k_2}; \IEEEyessubnumber\\
	\left.\frac{\partial G_s}{\partial y_2}\right|_{y_2 = 0} 
	&=& \left.\frac{\partial G_s}{\partial y_2}\right|_{y_2 = 1} 
	    \e^{\i k_2}. \IEEEyessubnumber
\end{IEEEeqnarray}

Eliminating the first-order term in (\ref{equ:WaveEquation}) by the
transformation $G_s = \bar{G}_s \e^{\i k M y_1 / \beta^2}$, one obtains
\begin{equation}
    \beta^2 \frac{\partial^2 \bar{G}_s}{\partial y_1^2} 
    + \frac{\partial^2 \bar{G}_s}{\partial y_2^2} 
    + \frac{\partial^2 \bar{G}_s}{\partial y_3^2} 
    + \left(\frac{k}{\beta}\right)^2 \bar{G}_s = 0,
    \label{equ:StretchedWaveEquation}
\end{equation}
Note here we can use either the non-orthogonal transformation commonly used in
previous works or the simple stretching transformation shown in
section~\ref{sec:GreensFunction}. The two would yield identical results.
However since section~\ref{sec:GreensFunction} uses the latter, here we choose
to use the former just for comparison. Introducing the non-orthogonal
coordinate transformation $\bxi_1 = (y_1 - h F(y_2)) / \beta$, $\bxi_2 = y_2$ and
$\bxi_3 = y_3$ yields,
\begin{equation}
    \frac{\partial^2 \bar{G}_s}{\partial \bxi_1^2} %
    + \frac{\partial^2 \bar{G}_s}{\partial \bxi_2^2}%
    + \frac{\partial^2 \bar{G}_s}{\partial \bxi_3^2}%
    - 2 \bar{h} F^\prime(\bxi_2) \frac{\partial^2 \bar{G}_s}{\partial \bxi_1
    \partial \bxi_2}%
    - \bar{h} F^{\prime\prime}(\bxi_2) \frac{\partial \bar{G}_s}{\partial \bxi_1}%
    + \bar{h}^2 F^{\prime 2}(\bxi_2) \frac{\partial^2 \bar{G}_s}{\partial \bxi_1^2}%
    + \bar{k}^2 \bar{G}_s = 0,
    \label{equ:variableChangedEquation}
\end{equation}
where the stretched constants are defined as $\bar{h} = h/\beta$ and $\bar{k} =
k/\beta$. Now the boundary conditions read
\begin{IEEEeqnarray}{rCl}
    \label{BCAfterChangeVariable}
    \left.\frac{\partial \bar{G}_s}{\partial \bxi_3}\right|_{\bxi_3=0} 
	&=&  i k_3 \e^{-\i k_1 (\bxi_1 + \bar{h} F(\bxi_2))} 
	\e^{-\i k_2 \bxi_2 }  \quad \bxi_1 < 0;\IEEEyesnumber \IEEEyessubnumber\\
	\bar{G}_s|_{\bxi_3 = 0} 
	&=& 0, \quad \bxi_1 > 0; \IEEEyessubnumber\\
	\bar{G}_s|_{\bxi_2 = 0} 
	&=& \bar{G}_s|_{\bxi_2 = 1} \e^{\i k_2}; \IEEEyessubnumber\\
	\left.\frac{\partial \bar{G}_s}{\partial \bxi_2}\right|_{\bxi_2 = 0} 
	&=& \left.\frac{\partial \bar{G}_s}{\partial \bxi_2}\right|_{\bxi_2 = 1}
	    \e^{\i k_2}. \IEEEyessubnumber
\end{IEEEeqnarray}

We can now perform the Fourier transform along the $\bxi_1$ direction, i.e.
\begin{equation}
    \mathcal{G}(s, \bxi_2, \bxi_3) = \int_{-\infty}^{\infty} \bar{G}_s(\bxi_1, \bxi_2, \bxi_3) \e^{\i s \bxi_1} \ud \bxi_1,
\end{equation}
(\ref{equ:variableChangedEquation}) then reduces to 
\begin{equation}
    \frac{\partial^2 \mathcal{G}}{\partial \bxi_2^2}%
    + \frac{\partial^2 \mathcal{G}}{\partial \bxi_3^2}%
    + 2 \i s \bar{h} F^\prime(\bxi_2) \frac{\partial \mathcal{G}}
    { \partial \bxi_2}%
    + \i s\bar{h} F^{\prime\prime}(\bxi_2) \mathcal{G}%
    - s^2 \bar{h}^2 F^{\prime 2}(\bxi_2) \mathcal{G}%
    + (\bar{k}^2-s^2) \mathcal{G} = 0.
    \label{equ:FourieredEquation}
\end{equation}
Upon use is made of the last two boundary conditions shown in
(\ref{BCAfterChangeVariable}), (\ref{equ:FourieredEquation}) can be solved by
using the method of separation variables, such that for $\bxi_3>0$ \add{(the
corresponding result for $\bxi_3<0$ is similar due to the antisymmetry)}
\begin{equation}
    \mathcal{G}(s, \bxi_2, \bxi_3) = \sum_{n = -\infty}^{\infty} A_n(s)
    \e^{-\gamma_n \bxi_3} \e^{-\i s \bar{h} F(\bxi_2)} \e^{\i \chi_n \bxi_2},
    \label{equ:determinedForm}
\end{equation}
where $\chi_n = 2n\pi - k_2$, $\gamma_n = \sqrt{s^2 - \kappa_n^2}$, and
$\kappa_n = \sqrt{\bar{k}^2 - \chi_n^2}$. The complex function $A_n(s)$ will
need to be determined by making use of the first two boundary conditions shown
in (\ref{BCAfterChangeVariable}) by using the Wiener-Hopf method,
i.e.
\begin{IEEEeqnarray}{rCl}
   \mathcal{G}^\prime(s, \bxi_2,  0) 
   &=& \frac{k_3}{(s - k_1)} 
   \sum_{n=-\infty}^{\infty} 
   E_n(s)\e^{-i s \bar{h} F(\bxi_2)} \e^{\i \chi_n \bxi_2} 
   + \sum_{-\infty}^{\infty}
   \mathcal{G}_n^{\prime+}(s)\e^{-i s \bar{h} F(\bxi_2)} \e^{\i \chi_n \bxi_2};
   \IEEEyesnumber\IEEEyessubnumber \IEEEyessubnumber\\
   \mathcal{G}(s,\bxi_2, 0) 
   &=& \sum_{n=-\infty}^{\infty}
   \mathcal{G}_{n}^-(s) \e^{-i s \bar{h} F(\bxi_2)} \e^{\i \chi_n \bxi_2}.
  \IEEEyessubnumber 
\end{IEEEeqnarray}   
where the symbol $\prime$ denotes the first derivative with respect to $\bxi_3$,
and $\mathcal{G}_n^-(s)$ and $\mathcal{G}_n^{\prime+}(s)$ are the expansion
coefficients of functions $\mathcal{G}^-(s, \bxi_2, 0)$ and $\mathcal{G}^{\prime
+}(s, \bxi_2, 0)$ using the basis functions $\e^{-\i s \bar{h} F(\bxi_2)} \e^{\i
\chi_n \bxi_2}, n =0, \pm 1, \pm 2 ...$, and they are unknown at this stage.
$E_n$ can be found to be 
\begin{equation}
    E_n(s) = \int_0^1 \e^{\i (k-s)\bar{h}F(\bxi_2)}\e^{-\i 2 n \pi \bxi_2} \ud \bxi_2
\end{equation}

Making use of orthogonality of the basis functions $\e^{-\i s \bar{h} F(\bxi_2)}
\e^{\i \chi_n \bxi_2}, n =0, \pm 1, \pm 2 ...$, we arrive at the following
matching conditions for mode $n$, i.e.
\begin{IEEEeqnarray}{rCl}
 -\gamma_n A_n(s)
&=&\frac{k_3}{(s - k_1)} 
E_n(s) + \mathcal{G}^{\prime + }_n(s); \IEEEyesnumber\IEEEyessubnumber\\
A_n(s) 
&=& \mathcal{G}_{n}^-(s).\IEEEyessubnumber
\end{IEEEeqnarray}
We can proceed by eliminating $A_(s)$ and arrive at 
\begin{equation}
    \gamma_n \mathcal{G}^-_n(s) + \frac{k_3}{s-k_1} E_n(s) +
    \mathcal{G}_n^\prime (s) = 0.
    \label{equ:WienerHopfEquation}
\end{equation}
Again, when $E_n(s)$ is assumed to be a factor of both $\mathcal{G}_n^-(s)$ and
$\mathcal{G}_n^{\prime +}(s)$ \add{it becomes a routine procedure to factorize
the kernel as $\sqrt{s - \kappa_n}\sqrt{s + \kappa_n}$,} and 
\begin{equation}
    A(s) = \mathcal{G}^-_n(s) 
    = \frac{-k_3 E_n(s)}{(s-k_1)}  
    \frac{1}{\sqrt{s - \kappa_n}}
    \frac{1}{\sqrt{k_1 + \kappa_n}}.
\end{equation}
With the same definition of $r$ and $\theta$ shown in
section~\ref{sec:GreensFunction}, the scattered pressure field $G_s$ can be
found to be
\begin{equation}
    G_s(r, \theta, y) 
    = \e^{\i k M x / \beta^2}\sum_{n = -\infty}^{\infty}
    \frac{-k_3\e^{\i \chi_n y} }{\sqrt{k_1 + \kappa_n} }
        \left[
	\frac{1}{2\pi}
	\int_{-\infty}^{\infty}
	\frac{E_n(s)}{(s-k_1)}  
	\frac{1}{\sqrt{s - \kappa_n}}
	\e^{(-i s  \cos \theta  - \gamma_n  \sin\theta) r} \ud s  
    \right],
    \label{equ:finalIntegralEquationAnotherDecomp}
\end{equation}
where the integral is along the path $P_0$ shown in
figure~\ref{fig:integralPath}.

Comparing (\ref{equ:GsIntegral}) and
(\ref{equ:finalIntegralEquationAnotherDecomp}) we see that when $n=0$ the
summands in these two equations are equal to each other. However, for other
values of $n$, the summands differ. This already signals a problem in the
validity of the assumption that $E_n(s)$ is a factor in $\mathcal{R}_n^-$ and
$\mathcal{R}_n^{\prime+}$, i.e. if such an assumption were true, the two
methods should yield completely identical results.
%\begin{equation} G(\boldsymbol{x}; \boldsymbol{y}, \omega) =
%A(\boldsymbol{x})\e^{\i\frac{kM}{\beta^2}y_1 } \e^{-\i k_2 y_2} \left\{
%\e^{-\i (k_1 y_1 /\beta + k_3 y_3)} + \sum_{n = -\infty}^{\infty} C_n(r,
%\theta) \frac{-k_3}{\sqrt{k_1 + \kappa_n}} \e^{\i 2 n \pi y_2}\right\}
%\end{equation} 
\bibliography{cleanRef} \bibliographystyle{apalike}

\end{document}